\documentclass[fleqn,usenatbib,usedcolumn]{mnras}

\usepackage[british]{babel}             % British English hyphenation
\usepackage{newtxtext}                  % Good fonts
   \usepackage[slantedGreek]{newtxmath}    %   "    "   (slanted Greek)
   \usepackage[T1]{fontenc}                % Good font encoding
   \usepackage{graphicx}                   % Including Figures
\usepackage{lscape}
\usepackage{longtable}
\usepackage{tabu}
\usepackage{url}
\usepackage{breakurl}
\renewcommand{\vec}[1]{\boldsymbol{#1}}
\title[Modelling \textit{Planck} clusters]{Physical modelling of galaxy clusters detected by the \textit{Planck} satellite}

\author[K. Javid et al.]
{Kamran Javid$^{1, 2}$\thanks{E-mail: kj316@mrao.cam.ac.uk},
Malak Olamaie$^{3,1}$,
Yvette C. Perrott$^{1}$,
Pedro Carvalho$^{1}$,
\newauthor
Keith J. B. Grainge$^{4}$,
Michael P. Hobson$^{1}$,
Clare Rumsey$^{1}$,
and Richard D. E. Saunders$^{1,2}$ 
\\
$^{1}$Astrophysics Group, Cavendish Laboratory, JJ Thomson Avenue, Cambridge CB3 0HE, UK\\
$^{2}$Kavli Institute for Cosmology Cambridge, Madingley Road, Cambridge, CB3 0HA, UK\\
$^{3}$Imperial Centre for Inference and Cosmology (ICIC), Imperial College, Prince Concort Road, London, SW7 2AZ, UK\\
$^{4}$Jodrell Bank Centre for Astrophysics, School of Physics and Astronomy, The University of Manchester, M13 9PL, UK
}

\date{Accepted XXX. Received YYY; in original form ZZZ}

\pubyear{2018}

\begin{document}
\label{firstpage}
\pagerange{\pageref{firstpage}--\pageref{lastpage}}
\maketitle

\begin{abstract}
We present a comparison of mass estimates for $54$ galaxy cluster candidates from the second \textit{Planck} catalogue (PSZ2) of Sunyaev--Zel'dovich sources. We compare the mass values obtained with data taken from the Arcminute Microkelvin Imager (AMI) radio interferometer system and from the \textit{Planck} satellite. The former of these uses a Bayesian analysis pipeline that parameterises a cluster in terms of its physical quantities, and models the dark matter and baryonic components of a cluster using Navarro-Frenk-White (NFW) and generalised-NFW profiles respectively. Our mass estimates derived from \textit{Planck} data are obtained from the results of the Bayesian detection algorithm PowellSnakes, are based on the methodology detailed in the PSZ2 paper, and produce two sets of mass estimates; one estimate is calculated directly from the angular radius $\theta$ -- integrated Comptonisation parameter $Y$ posterior distributions, and the other uses a `slicing function' to provide information on $\theta$ based on X-ray measurements and previous \textit{Planck} mission samples. 
We find that for $37$ of the clusters, the AMI mass estimates are lower than both values obtained from \textit{Planck} data. However the AMI and slicing function estimates are within one combined standard deviation of each other for $31$ clusters. %\\mnras only allows one paragraph for abstract.
We also generate cluster simulations based on the slicing-function mass estimates, and analyse them in the same way as we did the real AMI data. We find that inclusion in the simulations of radio-source confusion, CMB noise and measurable radio-sources causes AMI mass estimates to be systematically low.
\end{abstract}


%%%%%%%%%%%%%%%%%%%%%%%%%%%%%%%%%%%%%%%%%%%%%%%%%%%%%%%%%%%%%%%%%%%%%
%%%%%%%%%%%%%%%%%%%%%%%%%%%%%%%%%%%%%%%%%%%%%%%%%%%%%%%%%%%%%%%%%%%%%
%%%%%%%%%%%%%%%%%%%%%%%%%%%%%%%%%%%%%%%%%%%%%%%%%%%%%%%%%%%%%%%%%%%%%

\section{Introduction}

%Galaxy clusters are the most massive gravitationally bound known objects in the Universe and thus are critical tracers of the formation of large-scale structure. The abundance of galaxy clusters at different redshifts and mass are therefore excellent cosmological probes of both the early and modern Universe, giving one the opportunity to deduce values for the Hubble parameter, $H_{0}$ \citep{1978obco.meet....1G}, the mean matter density, $\Omega_{\rm M}$, and many other cosmological parameters (see \citealt{2006JCAP...10..014S} for example). 

In the local Universe and out to redshifts of around two, clusters of galaxies are observed as massive gravitationally bound structures, often roughly spherical and with very dense central cores.
It is over eighty years ago that it was first postulated that a galaxy cluster's mass is dominated by dark matter (%\textcolor{red}{\citealt{1959HDP....53..390Z}} 
\citealt{1933AcHPh...6..110Z} and \citealt{1937ApJ....86..217Z}). More recently it has been shown that dark matter contributes $\approx90\%$ of the cluster mass (see e.g. \citealt{2006ApJ...640..691V} and \citealt{2011ApJS..192...18K}). Stars, gas and dust in galaxies, as well as a hot ionised intra-cluster medium (ICM), make up the rest of the mass in a cluster, with the latter being the most massive baryonic component. The galaxies emit in the optical and infrared wavebands, whilst the ICM emits in X-ray via thermal Bremsstrahlung and also interacts with cosmic microwave background (CMB) photons via inverse Compton scattering. This last effect is what is known as the Sunyaev--Zel'dovich (SZ) effect \citep{1970CoASP...2...66S}.
It is this effect which is detected by the \textit{Planck} satellite and the Arcminute Microkelvin Imager (AMI) radio interferometer system, which are the telescopes featured in this analysis. The clusters detected by \textit{Planck} form the basis of the sample considered in this work. \citet{2015A&A...580A..95P} (from here on YP15) present the results of the AMI follow-up of \textit{Planck} clusters-- this follow-up is analysed using the `observational model' \citep{2012MNRAS.421.1136A}, which parameterises a cluster in terms of its integrated Comptonisation parameter $Y$ and angular scale $\theta$. YP15 find that these AMI estimates for $Y$ are consistently lower than the values obtained from \textit{Planck} data, and conclude that this may indicate that the cluster pressure profiles are deviating from the `universal' one. Here, we try to overcome this by considering a model which uses redshift information to break this degeneracy. %By fitting simple parametric cluster models to the observed dataset, one would like to make model-dependent inferences about the cluster's physical parameters, i.e. to calculate the probability distribution of these parameters. This task is most conveniently carried out through Bayesian inference. 
We use a physical model largely based on the one described in \citet{2012MNRAS.423.1534O} (from here on MO12), with data obtained from AMI of clusters detected by \textit{Planck} (including ones which were detected after the analysis in YP15 was carried out). 
We also consider the cluster mass estimates given in the \textit{Planck} cluster catalogue \citet{2016A&A...594A..27P} and compare them with the values obtained using AMI data.
Furthermore we use the \textit{Planck} cluster catalogue mass estimates as inputs to simulations which are then analysed in the same way as real AMI observations. 

In Section~\ref{sec:telescopes} we give an overview of the \textit{Planck} mission and AMI in the context of \textit{Planck} observed clusters. In Section~\ref{sec:models} we review how the physical modelling process works with data obtained from AMI, and we summarise the methodology used to obtain the mass estimates given in \citet{2016A&A...594A..27P}. Sections~\ref{sec:results_i} and~\ref{sec:results_ii} present the results of our analysis, including simulated AMI data which used mass estimates obtained from \textit{Planck} data as inputs.

A `concordance' flat $\Lambda$CDM cosmology is assumed: $\Omega_{\rm M} = 0.3$, $\Omega_{\Lambda} = 0.7$, $\Omega_{\rm R} = 0$, $\Omega_{\rm K} = 0$, $h=0.7$, $H_{0} = 100~h~\rm km~s^{-1}$~Mpc$^{-1}$, $\sigma_{8} = 0.8$, $w_{0} = -1$, and $w_{\rm a} = 0$. The first four parameters correspond to the (dark + baryonic) matter, the cosmological constant, the radiation, and the curvature densities respectively. $h$ is the dimensionless Hubble parameter, while $H_{0}$ is the Hubble parameter now and $\sigma_{8}$ is the power spectrum normalisation on the scale of $8$~$h^{-1}$~Mpc now. $w_{\rm 0}$ and $w_{\rm a}$ are the equation of state parameters of the Chevallier-Polarski-Linder parameterisation \citep{2001IJMPD..10..213C}.

%%%%%%%%%%%%%%%%%%%%%%%%%%%%%%%%%%%%%%%%%%%%%%%%%%%%%%%%%%%%%%%%%%%%%%%%%%%%%%%
%%%%%%%%%%%%%%%%%%%%%%%%%%%%%%%%%%%%%%%%%%%%%%%%%%%%%%%%%%%%%%%%%%%%%%%%%%%%%%%
%%%%%%%%%%%%%%%%%%%%%%%%%%%%%%%%%%%%%%%%%%%%%%%%%%%%%%%%%%%%%%%%%%%%%%%%%%%%%%%

\section{\textit{Planck} and AMI telescopes, and the cluster sample}
\label{sec:telescopes}

%%%%%%%%%%%%%%%%%%%%%%%%%%%%%%%%%%%%%%%%%%%%%%%%%%%%%%%%%%%%%%%%%%%%%%%%%%%%%%%
%%%%%%%%%%%%%%%%%%%%%%%%%%%%%%%%%%%%%%%%%%%%%%%%%%%%%%%%%%%%%%%%%%%%%%%%%%%%%%%

\subsection{\textit{Planck} mission}
%The \textit{Planck} mission\footnote{\url{http://www.esa.int/\textit{Planck}/}} is a European Space Agency (ESA) mission, whose primary objective is to investigate the CMB, the relic radiation from the origin of our Universe. The \textit{Planck} telescope was a space telescope which operated using a low frequency instrument and a high frequency instrument, which had a combined frequency range 30 - 857~GHz.
The combination of the \textit{Planck} satellite's low frequency and high frequency instruments provide nine frequency channels in the range 37~GHz -- 857~GHz.
Of particular importance for the work described here are the \textit{Planck} cluster-catalogues (see \citealt{2014A&A...571A..29P}, \citealt{2015A&A...581A..14P} and \citealt{2016A&A...594A..27P} for papers relating to catalogues PSZ1, PSZ1.2 and PSZ2 respectively, where `PSZX' refers to the X\textsuperscript{th} \textit{Planck} SZ catalogue). These provide, for example, cluster candidate positions, redshift ($z$) values, integrated Comptonisation parameter values and mass ($M$) estimates. PSZ2 is the most recent all-sky \textit{Planck} cluster catalogue, and is the one which we refer to in this paper unless stated otherwise.

%%%%%%%%%%%%%%%%%%%%%%%%%%%%%%%%%%%%%%%%%%%%%%%%%%%%%%%%%%%%%%%%%%%%%%%%%%%%%%%
%%%%%%%%%%%%%%%%%%%%%%%%%%%%%%%%%%%%%%%%%%%%%%%%%%%%%%%%%%%%%%%%%%%%%%%%%%%%%%%

\subsection{PSZ2 redshift values}
%The \textit{Planck} satellite was not intended to provide redshift measurements itself, and so the values published in the \textit{Planck} catalogues were obtained from telescopes which operate in the optical/infrared or X-ray wavebands. 
Catalogue $z$ values are measured in the optical / infrared or X-ray, with major input from the Sloan Digital Sky Survey \citep{2000AJ....120.1579Y}. A number of cluster catalogues have been extracted from these data (see e.g. \citealt{2010ApJS..191..254H}, \citealt{2012ApJS..199...34W}, and \citealt{2014ApJ...785..104R}), providing estimates of both spectroscopic and photometric $z$ values, the reliability of the latter values falls as $z$ increases. In the X-ray part of the spectrum, the Meta-Catalogue of X-ray detected Clusters of galaxies, or MCXC \citep{2011A&A...534A.109P} has a substantial number of matches with the \textit{Planck}-catalogue clusters. The MCXC is from the available catalogues based on the ROSAT All-Sky Survey \citep{1999A&A...349..389V} as well as serendipitous X-ray catalogues (see e.g. \citealt{1990ApJS...72..567G}). MCXC contains only clusters with measured $z$, but does not state the redshift type or source.
%Other sources of X-ray redshifts come from the XMM Newton telescope \citep{2007A&A...469..363B} and the Chandra telescope \citep{2002PASP..114....1W}.  
Further sources of \textit{Planck} catalogue clusters candidate $z$s are the Russian-Turkish Telescope \citep{2015A&A...582A..29P} and the ENO telescopes in the Canary Islands \citep{2016A&A...586A.139P}; for each $z$ these state whether it was obtained photometrically or spectroscopically. 
%The technique used to measure the redshift (i.e. spectroscopic or photometric) for the final cluster sample considered is given in the results Table (see appendix \ref{sec:results_Table}).

%%%%%%%%%%%%%%%%%%%%%%%%%%%%%%%%%%%%%%%%%%%%%%%%%%%%%%%%%%%%%%%%%%%%%%%%%%%%%%%
%%%%%%%%%%%%%%%%%%%%%%%%%%%%%%%%%%%%%%%%%%%%%%%%%%%%%%%%%%%%%%%%%%%%%%%%%%%%%%%

\subsection{AMI}

AMI is an interferometer system near Cambridge, UK, designed for SZ studies (see e.g. \citealt{2008MNRAS.391.1545Z}). It consists of two arrays: the Small Array (SA), optimised to couple to SZ signal, with an angular resolution of $\approx 3$~arcmin and sensitivity to structures up to $\approx 10$~arcmin in scale; and the Large Array (LA), with angular resolution of $\approx 30$~arcsec, which is largely insensitive to SZ, and is used to characterise and subtract confusing radio-sources.  Both arrays operate at a central frequency of $\approx 15.7$~GHz, and at the time the AMI data for this paper were taken, with a bandwidth of $\approx 4.3\,$~GHz, divided into six channels. A summary of AMI's characteristics is given in Table~\ref{tab:ami}. Note that AMI has recently received a new digital correlator \citep{2018MNRAS.475.5677H}, but all real data used in this work were obtained by the system with its old analogue correlator. 

Our pointing strategy for each cluster was as follows. Clusters were observed using a single pointing centre on the SA, which has a primary beam of size $\approx 20~$arcmin FWHM, to noise levels of $\lessapprox 120~\mu \rm{Jy}~\rm{beam}^{-1}$. To cover the same area with the LA, which has a primary beam of size $\approx $ six~arcmin FWHM, the cluster field was observed as a 61-point hexagonal raster. The noise level of the raster was $\lessapprox 100~\mu \rm{Jy}~\rm{beam}^{-1}$ in the central 19 pointings, and slightly higher in the outer regions. The observations for a given cluster field were carried out simultaneously on both arrays, and the average observation time per cluster was $\approx 30~$ hours.
%The cluster observations, data calibration and data flagging were carried out as described in Section~3 of YP15. 
The observations were carried out between 2013 and 2015, and so they began before PSZ2 was published. This means that the AMI pointing centre coordinates in general were not the same as those published in the final \textit{Planck} catalogue which was released in 2015. This is discussed in the context of the cluster centre offset parameters in Section~\ref{subsub:clus_priors}. 

Data from both arrays were flagged for interference and calibrated using the AMI in-house software package \textsc{REDUCE}. Flux calibration was applied using contemporaneous observations of the primary calibration sources 3C 286, 3C 48, and 3C 147. The assumed flux densities for 3C 286 were converted from Very Large Array total-intensity measurements \citep{2013ApJS..204...19P} and are consistent with \citet{1987Icar...71..159R}. % model of Mars transferred onto an absolute scale, using results from the Wilkinson Microwave Anisotropy Probe. 
The assumed flux densities for 3C 48 and 3C 147 were based on long-term monitoring with the SA using 3C 286 for flux calibration. Phase calibration was applied using interleaved observations of a nearby bright source selected from the Very Long Baseline Array Calibrator survey \citep{2008AJ....136..580P}; in the case of the LA, a secondary amplitude calibration was also applied using observations of the phase calibration source on the SA.

%\footnote{\url{https://www.mrao.cam.ac.uk/research/arcminute-microkelvin-imager-ami-digital-correlator/}}. %All of the data obtained in this analysis was done using the analogue correlator. %, but in future work a selection of clusters detected by \textit{Planck} will be re-analysed with the new digital correlator system.

\begin{table}
\centering
\begin{tabular}{{l}{c}{c}}
\hline
 & SA & LA \\
\hline 
Antenna diameter & $3.7~\rm{m}$ & $12.8~\rm{m}$ \\
Number of antennas & $10$ & $8$ \\
Baseline lengths (current) & $5-20~\rm{m}$ & $18-110~\rm{m}$ \\
Primary beam FWHM (at $15.7~\rm{GHz}$) & $20.1~\rm{arcmin}$ & $5.5~\rm{arcmin}$ \\
Typical synthesised beam FWHM & $3~\rm{arcmin}$ & $30~\rm{arcsec}$ \\
Flux sensitivity & $30~\rm{mJy}~\rm{s}^{1/2}$ & $3~\rm{mJy}~\rm{s}^{1/2}$ \\
\hline
\end{tabular}
\caption{Summary of AMI characteristics.}\label{tab:ami}
\end{table}

%%%%%%%%%%%%%%%%%%%%%%%%%%%%%%%%%%%%%%%%%%%%%%%%%%%%%%%%%%%%%%%%%%%%%%%%%%%%%%%
%%%%%%%%%%%%%%%%%%%%%%%%%%%%%%%%%%%%%%%%%%%%%%%%%%%%%%%%%%%%%%%%%%%%%%%%%%%%%%%

\subsection{Selection of the cluster sample}
%The second \textit{Planck} catalogue of clusters \citep{2016A&A...594A..27P} 
PSZ2 contains 1653 cluster candidates detected in the all-sky 29 month mission. The initial cluster selection criteria for AMI closely resembles that described in YP15, with a few modifications as follows. 
\begin{itemize}
\item The lower $z$ limit $ z \leq 0.100 $ was relaxed here, to see how well AMI data can constrain the the physical model parameters at low redshift. However it is important to realise that the sample at $z \leq 0.100$ were not observed specifically for the purpose of this work, but were part of other observation projects. %as AMI did not observe all the \textit{Planck} clusters which it was capable of at this $z$- a select few clusters were observed in aid of other projects which included these clusters.
\item The \textit{Planck} signal-to-noise ratio (S/N) lower bound was reduced to $4.5$.
\item The automatic radio-source environment rejection remained the same. However the manual rejection was done on a map-by-map basis-- see Section~\ref{sec:results_i}.
\item Note that the observation declination limits $20^{\circ} < \delta < 87^{\circ}$ were kept. %Note that with the new digital correlator, AMI is capable of observing at lower declinations without suffering from excessive interference from geostationary satellites, and so observations of \textit{Planck} clusters at lower $\delta$ will be carried out in future work (CHECK WITH YVETTE).
\end{itemize}
This led to an initial sample size of 199 clusters, %which had been detected by \textit{Planck} and re-observed with AMI to produce data which could be run through the data analysis pipeline. 
The maximum and minimum values of some key parameters for this sample from the \textit{Planck} catalogue are given in Table~\ref{tab:initial_sample}. Note that $M_{\rm{SZ}}$ is taken in PSZ2 as the hydrostatic equilibrium mass $M(r_{500})$, assuming the best-fit $Y-M$ relation (see Section~\ref{subsubsec:planckmass}).%Note that $M_{\rm{SZ}}$ is taken in \citet{2016A&A...594A..27P} as the hydrostatic equilibrium mass, $M(r_{500})$, assuming the best-fit $Y-M$ relation (see Section~\ref{sec:models}). %This will be elaborated on in detail in Section~\ref{sec:models} when the \textit{Planck} scaling relation methodology is described in detail. But for now it suffices to know that this represents the total enclosed mass up to a radius $r_{500}$.
\begin{table}
\centering
\begin{tabular}{{l}{c}{c}}
\hline
Parameter & Minimum value & Maximum value \\
\hline 
Declination & $20.31^{\circ}$ & $86.24^{\circ}$ \\
$z$ & $0.045$ & $0.83$ \\
S/N & 4.50 & 28.40 \\
$M_{\rm{SZ}}$~($\times 10^{14}~M_{\mathrm{Sun}}$) & $1.83$ & $10.80$ \\
\hline
\end{tabular}
\caption{Minimum and maximum values for a selection of parameters taken from PSZ2 for the AMI sample of 199 clusters.}\label{tab:initial_sample}
\end{table}

%The method by which the observations were made, making use of the AMI small array (SA) to target the actual objects of interest, and the large array (LA) to observe the contaminating radio-sources present in the field of observation, was as described in YP15. 

%%%%%%%%%%%%%%%%%%%%%%%%%%%%%%%%%%%%%%%%%%%%%%%%%%%%%%%%%%%%%%%%%%%%%%%%%%%%%%%
%%%%%%%%%%%%%%%%%%%%%%%%%%%%%%%%%%%%%%%%%%%%%%%%%%%%%%%%%%%%%%%%%%%%%%%%%%%%%%%
%%%%%%%%%%%%%%%%%%%%%%%%%%%%%%%%%%%%%%%%%%%%%%%%%%%%%%%%%%%%%%%%%%%%%%%%%%%%%%%

\section{AMI data analysis and PSZ2 scaling relations methodology}
\label{sec:models}
Our AMI Bayesian data analysis pipeline, \textsc{McAdam} closely resembles the one described in \citet{2009MNRAS.398.2049F} (FF09 from here on). In this Section the key aspects of the analysis are summarised, and also we note modifications specific to the work of this paper.

%%%%%%%%%%%%%%%%%%%%%%%%%%%%%%%%%%%%%%%%%%%%%%%%%%%%%%%%%%%%%%%%%%%%%%%%%%%%%%%
%%%%%%%%%%%%%%%%%%%%%%%%%%%%%%%%%%%%%%%%%%%%%%%%%%%%%%%%%%%%%%%%%%%%%%%%%%%%%%%

\subsection{A physical model for AMI data}
\label{subsec:phys}
In the implementation of McAdam used here, we in large employed the model of MO12 to derive physical properties of a galaxy cluster (i.e. mass, pressure, density, radius and temperature values) from the data obtained from an SZ-detecting interferometer plus $z$-information.
The model assumes an Navarro-Frenk-White (NFW) profile \citep{1995MNRAS.275..720N} for the dark matter density as a function of cluster radius $r$,%the dark matter component of a galaxy cluster
\begin{equation}\label{eqn:nfw}
\rho_{\rm dm}(r) = \frac{\rho_{\rm s}}{\left(\frac{r}{r_{\rm s}}\right)\left(1+\frac{r}{r_{\rm s}}\right)^{2}},
\end{equation}
%$\rho_{\rm dm}(r)$ is 
where $\rho_{\rm s}$ is an overall density normalisation coefficient and $r_{\rm s}$ is a characteristic radius defined by $r_{\rm s} \equiv r_{200}/c_{200}$ and is the radius at which the logarithmic slope of the profile $ \mathrm{d }\ln \rho(r) / \mathrm{d }\ln r$ is $-2$. $r_{200}$ is the radius at which the \textit{mean} cluster density is $200 \times \rho_{\rm crit}(z)$. $\rho_{\rm crit}(z)$ is the critical density of the Universe at the cluster $z$ which is given by $\rho_{\rm crit}(z) = 3H(z)^{2}/8\pi G$ where $H(z)$ is the Hubble parameter (at the cluster redshift) and $G$ is Newton's constant. 
$c_{200}$ is the concentration parameter at this radius. Following \citet{2013MNRAS.430.1344O}, we calculate $c_{200}$ for an NFW dark matter density profile taken from the expression in \citet{2009MNRAS.393.1235C}
\begin{equation}\label{eqn:c200}
c_{200} = \frac{5.26}{1+z} \left( \frac{M(r_{200})}{10^{14}h^{-1}M_{\mathrm{Sun}}} \right)^{-0.1}.
\end{equation}
The $1/(1+z)$ factor comes from \citet{2001astro.ph.11069W} and is obtained from N-body simulated dark matter halos between $z=0$ and $z=7$. The remainder of the relation was derived in \citet{2007MNRAS.381.1450N} by fitting a power-law for $c_{200}$ to N-body simulated cluster data. Note that the sample used in \citet{2007MNRAS.381.1450N} was assumed to contain clusters that are relaxed.
%The relation is derived by fitting a power-law to N-body simulations of clusters which are assumed to be relaxed.
In equation~\ref{eqn:c200} $M(r_{200})$ is the mass enclosed at radius $r_{200}$. Thus for given values of $z$ and $M(r_{200})$ (which are input parameters to the model, see Section~\ref{subsub:clus_priors}), $c_{200}$ can be calculated. Furthermore if we take $M(r_{200}) = 200 \times \frac{4\pi}{3} \rho_{\rm crit}(z) r_{200}^{3}$ then we can also calculate $r_{200}$ and so $r_{\rm s}$.

%\citet{1996ApJ...462..563N} showed through N-body simulations that this model works well for spherically averaged dark matter halos, as illustrated in Figure 2 of \citet{1997ApJ...490..493N}.

%Although dark matter forms the major component of a cluster's mass, careful consideration of the baryonic matter is vital in order to obtain a realistic model. %Among the first attempts to model the gas content of cluster was the work of \citet{1986MNRAS.222..323K}, who relied on self-similarity in clusters to model their properties as they evolved over time. More recently similar work was carried out with high-resolution cosmological simulations in \citet{2005ApJ...625..588K}, and was used to derive cluster parameters from Chandra observed clusters in \citet{2006ApJ...650..128K, 2006astro.ph..8330V, 2007ApJ...668....1N} and \citet{2007hvcg.conf..358N}. 
Following \citet{2007ApJ...668....1N}, the generalised NFW (GNFW) model is used to parameterise the electron pressure as a function of radius from the cluster centre
\begin{equation}\label{eqn:gnfw}
P_{\rm e}(r) = \frac{P_{\rm ei}}{\left(\frac{r}{r_{\rm p}}\right)^{c}\left(1+\left(\frac{r}{r_{\rm p}}\right)^{a}\right)^{(b-c)/a}},
\end{equation}
where $P_{\rm ei}$ is an overall pressure normalisation factor and $r_{\rm p}$ is another characteristic radius, defined by $r_{\rm p} \equiv r_{500}/c_{500}$ ($r_{500}$ is the radius at which the cluster density is $500 \times \rho_{\rm crit}(z)$). The parameters $a$, $b$ and $c$ describe the slope of the pressure profile at $ r / r_{\rm p} \approx 1$, $r / r_{\rm p} \gg 1 $ and $r / r_{\rm p} \ll 1$ respectively. For values $r / r_{\rm p} \ll 1$ the logarithmic slope ($ \mathrm{d} \ln P_{\mathrm{e}}(r) / \mathrm{d} \ln r $) converges to $-c$. For values $r / r_{\rm p} \gg 1$ the logarithmic slope converges to $-b$. The value of $a$ dictates how quickly (in terms of $r$) the slope switches between these two values, and in the limit that $a$ tends to zero, the logarithmic slope is $-(b+c)/2$ for all $r$. Consistent with many of the \textit{Planck} follow-up papers (see e.g. \citealt{2011A&A...536A...8P}) and with MO12 the slope parameters are taken to be $a = 1.0620$, $b=5.4807$ and $c = 0.3292$. These `universal' values are from \citet{2010A&A...517A..92A} and are the GNFW slope parameters derived for the standard self-similar case using scaling relations from a REXCESS sub-sample (of 20 well-studied low-$z$ clusters observed with XMM-Newton), as described in Appendix B of the paper \citep{2007A&A...469..363B}. We also use the Arnaud et al. value for the concentration parameter $c_{500} \equiv r_{500} / r_{\rm p}$ of $1.156$. 
We note however that in YP15 using simulations it was shown that the disagreement between \textit{Planck} and AMI parameter estimates may indicate pressure profiles deviating from the `universal' profile. %We note however that in YP15 from simulations it was found that the pressure profiles of some clusters (particularly ones with large angular scales: $\theta_{\rm{s}} = r_{\rm{s}} / D_{\rm{A}} \geq 5 $~arcmin, where $D_{A}$ is the angular diameter distance) may deviate from the `universal' profile.
%P and r don't appear in math font for some reason
For any model it is important to know the underlying assumptions which allow it to be valid. The four relevant assumptions in MO12 are
\begin{itemize}
\item The cluster is spherically symmetric.
\item The cluster is in hydrostatic equilibrium up to radius $r_{200}$. This means at any radius up to $r_{200}$ the outward pushing pressure force created by the pressure differential at that point must be equal to the gravitational binding force due to the mass enclosed within that radius (\citealt{1977ApJ...213L..99B}, see equation~4 of MO12). 
\item The gas mass fraction $f_{\rm gas}(r)$ is much less than unity up to radius $r_{200}$, so that the total mass is $M(r_{200}) \approx M_{\rm dm}(r_{200})$. Consequently the total mass out to $r_{200}$ is given by the integral of the dark matter density along the radius of the cluster (equation~5 of MO12). 
\item The cluster gas is assumed to be an ideal gas, so that its temperature can be trivially represented in terms of its pressure (equations~13 and~14 of MO12).
\end{itemize}

The calculation steps used for the present paper are as described in MO12 except for one modification. Previously, the mapping from $r_{200}$ to $r_{500}$ was a constant factor $\frac{2}{3}$ which was derived by solving the equation
%$F: r_{200} \rightarrow r_{500}$ %R doesn't like this notation
\begin{equation}\label{eqn:r200r5001}
\begin{split}
\left( \frac{r_{\rm s}}{r_{500}} \right)^{3} \left[ \ln \left( 1+\frac{r_{500}}{r_{\rm s}} \right) - \left( 1 + \frac{r_{\rm s}}{r_{500}} \right)^{-1} \right] = \frac{5}{2} \left( \frac{r_{\rm s}}{r_{200}} \right)^{3} \times \\ 
\left[ \ln \left( 1+\frac{r_{200}}{r_{\rm s}} \right) - \left( 1 + \frac{r_{\rm s}}{r_{200}} \right)^{-1} \right]
\end{split}
\end{equation}
for a range of values of $M(r_{200})$ and $z$. However, following \citet{2003ApJ...584..702H}, there is an analytic mapping from $r_{200}$ to $r_{500}$. Consider the equation
\begin{equation}\label{eqn:r200r5002}
g(r_{\rm s}/r_{500}) = \frac{5}{2} g(r_{\rm s}/r_{200}),
\end{equation}
where
\begin{equation}\label{eqn:r200r5003}
g(x) = x^{3} [ \ln (1 + x^{-1}) - (1 + x)^{-1} ].
\end{equation}
Equation \ref{eqn:r200r5002} requires that $g(r_{\rm s}/r_{500})$ be inverted so that
\begin{equation}\label{eqn:r200r5004}
\frac{r_{\rm s}}{r_{500}} = x \left( g_{500}=\frac{5}{2}f(r_{\rm s}/r_{200}) \right),
\end{equation}
where
\begin{equation}\label{eqn:r200r5005}
x(g_{500}) = \left[ a_{1}g_{500}^{2p} + \frac{9}{16} \right] ^{-1/2} + 2g_{500}.
\end{equation}
Here $p = a_{2} + a_{3} \ln g_{500} + a_{4}(\ln g_{500})^{2}$, and the four fitting parameters correspond to $a_{1} = 0.5116$, $a_{2} = -0.4283$, $a_{3} = -3.13 \times 10^{-3}$ and $a_{4} = -3.52 \times 10^{-5}$. This gives a fit to better than 0.3\% accuracy for $ 0 < c_{200} < 20 $ and is exact in the limit that $ c_{200} \rightarrow 0$. %Note that determining $r_{500}$ this way requires the assumption that $M(r_{500}) = M_{\rm dm}(r_{500})$. However this has already been implicitly assumed when we asserted that $M(r_{200}) = M_{\rm dm}(r_{200})$, since this means we make this assumption for all $r \leq r_{200}$, and $r_{500} < r_{200}$.

%%%%%%%%%%%%%%%%%%%%%%%%%%%%%%%%%%%%%%%%%%%%%%%%%%%%%%%%%%%%%%%%%%%%%%%%%%%%%%%
%%%%%%%%%%%%%%%%%%%%%%%%%%%%%%%%%%%%%%%%%%%%%%%%%%%%%%%%%%%%%%%%%%%%%%%%%%%%%%%

\subsection{Bayesian parameter estimation}

Given a model $\mathcal{M}$ and a data vector $\vec{\mathcal{D}}$, one can obtain model parameters (also known as input or sampling parameters) $\vec{\Theta}$ conditioned on $\mathcal{M}$ and $\vec{\mathcal{D}}$ using Bayes' theorem:
\begin{equation}\label{eqn:bayes}
Pr\left(\vec{\Theta}|\vec{\mathcal{D}},\mathcal{M}\right) = \frac{Pr\left(\vec{\mathcal{D}}|\vec{\Theta},\mathcal{M}\right)Pr\left(\vec{\Theta}|\mathcal{M}\right)}{Pr\left(\vec{\mathcal{D}}|\mathcal{M}\right)},
\end{equation}
where $Pr(\vec{\Theta}|\vec{\mathcal{D}},\mathcal{M}) \equiv \mathcal{P}(\vec{\Theta})$ is the posterior distribution of the model parameter set, $Pr(\vec{\mathcal{D}}|\vec{\Theta},\mathcal{M}) \equiv \mathcal{L}(\vec{\Theta})$ is the likelihood function for the data, $Pr(\vec{\Theta}|\mathcal{M}) \equiv \pi(\vec{\Theta})$ is the prior probability distribution for the model parameter set, and $Pr(\vec{\mathcal{D}}|\mathcal{M}) \equiv \mathcal{Z}(\vec{\mathcal{D}})$ is the Bayesian evidence of the data given a model $\mathcal{M}$. The evidence can be interpreted as the factor required to normalise the posterior over the model parameter space:
\begin{equation}\label{eqn:evidence}
\mathcal{Z}(\vec{\mathcal{D}}) = \int \mathcal{L}(\vec{\Theta}) \pi(\vec{\Theta})\, \mathrm{d}\vec{\Theta},
\end{equation} 
where the integral is carried out over the $N$-dimensional parameter space. Although $\mathcal{Z}(\vec{\mathcal{D}})$ is not important in the context of parameter estimation, it is central to the way that the posterior distributions are determined using the nested sampling algorithm \textsc{MultiNest} \citep{2009MNRAS.398.1601F}. \textsc{MultiNest} is a Monte Carlo algorithm which makes use of a transformation of the $N$-dimensional evidence integral into a much easier to evaluate one-dimensional integral, and generates samples from the posterior distribution $\mathcal{P}(\vec{\Theta})$ as a by-product.
The input parameters can be split into two subsets (which are assumed to be independent of one another): cluster parameters $\vec{\Theta}_{\rm cl}$ and radio-source or `nuisance' parameters $\vec{\Theta}_{\rm rs}$.

%%%%%%%%%%%%%%%%%%%%%%%%%%%%%%%%%%%%%%%%%%%%%%%%%%%%%%%%%%%%%%%%%%%%%%%%%%%%%%%

\subsubsection{Cluster parameter prior distributions}
\label{subsub:clus_priors}

Following FF09, the cluster parameters are assumed to be independent of one another, so that
\begin{equation}\label{eqn:cluspriors}
\pi\left(\vec{\Theta}_{\rm cl}\right) = \pi(M(r_{200}))\pi(f_{\rm gas}(r_{200}))\pi(z)\pi(x_{\rm c})\pi(y_{\rm c}).
\end{equation} 
$x_{\rm c}$ and $y_{\rm c}$ are the cluster centre offsets from the SA pointing centre, measured in arcseconds. The prior distributions assigned to the cluster parameters are the same as the ones used in \citet{2013MNRAS.430.1344O}, but with an alteration to the mass limits. Upon running \textsc{McAdam} on data from a few of the \textit{Planck} clusters, it was found that the posterior distributions of $M(r_{200})$ were hitting the lower bound $1 \times 10^{14}~M_{\mathrm{Sun}}$ used in \citet{2013MNRAS.430.1344O}. Hence for this analysis the lower limit on $M(r_{200})$ was decreased. Table~\ref{tab:clusterpriors} lists the type of prior used for each cluster parameter and the probability distribution parameters. Values for $z_{\rm Planck}$  were taken from the PSZ2 catalogue.%Note that allowing the cluster centre coordinates to vary partially accommodates for the issue that the AMI observations were not pointed at the most up to date coordinates published by \textit{Planck}, but it does not make up for the fact that the primary beam will still be offset from this point (CHECK WITH YVETTE).

\begin{table}
\centering
\begin{tabular}{{l}{c}}
\hline
Parameter & Prior distribution \\ 
\hline
$x_{\rm c}$ & $\mathcal{N}(0'', 60'')$ \\
$y_{\rm c}$ & $\mathcal{N}(0'', 60'')$ \\
$z$ & $\delta(z_{\rm Planck})$\\
$M(r_{200})$ & $\mathcal{U} [ \log (0.5\times 10^{14} M_{\rm{Sun}}), \log (50\times 10^{14} M_{\rm{Sun}})]$ \\
$f_{\rm gas}(r_{200})$ & $\mathcal{N}(0.13, 0.02)$ \\
\hline
\end{tabular}
\caption{Cluster parameter prior distributions. $\delta$ denotes a Dirac delta function, $\mathcal{U}$ is a uniform distribution and $\mathcal{N}$ is a normal distribution (parameterised by its mean and standard deviation).}\label{tab:clusterpriors}
\end{table}

%%%%%%%%%%%%%%%%%%%%%%%%%%%%%%%%%%%%%%%%%%%%%%%%%%%%%%%%%%%%%%%%%%%%%%%%%%%%%%%

\subsubsection{Measured radio-source parameter prior distributions}
\label{subsub:rs_priors}

Each radio-source recognised and measured by the LA can also be modelled in the analysis. Following FF09, each source can be parameterised by four variables: its position on the sky ($x_{\rm rs}$, $y_{\rm rs}$), its measured flux density $S_{\rm rs}$, and its spectral index $\alpha_{\rm rs}$. The latter of these quantities describes how the flux density of a radiating object depends on the frequency of the radiation. Assuming these are independent, then for source $i$ 
\begin{equation}\label{eqn:rspriors}
\pi(\vec{\Theta}_{\mathrm{rs, } i}) = \pi(x_{\mathrm{rs, } i})\pi(y_{\mathrm{rs, } i})\pi(S_{\mathrm{rs, } i})\pi(\alpha_{\mathrm{rs, } i}).
\end{equation} 
Delta functions are applied to the distributions on $x_{\rm rs}$ and $y_{\rm rs}$, due to the LA's ability to measure spatial positions to high accuracy: $\pi(x_{\rm rs}) = \delta(x_{\mathrm{rs, \, LA}})$, $\pi(y_{\rm rs}) = \delta(y_{\mathrm{rs, \, LA}})$. Delta priors were also set on $S_{\rm rs}$ and $\alpha_{\rm rs}$ (centred on the values measured by the LA), if the measured $S_{\rm rs}$ was less than four times the instrumental noise associated with the observation, and the source was more than 5 arcminutes away from the SA pointing centre: $\pi(S_{\rm rs}) = \delta(S_{\mathrm{rs, \, LA}})$, $\pi(\alpha_{\rm rs}) = \delta(\alpha_{\mathrm{rs, \, LA}})$. Otherwise, a Gaussian prior was set on $S_{\rm{rs}}$ centred at the LA measured value with a standard deviation equal to $40\%$ of the measured value ($\sigma_{\rm rs} = 0.4 \times S_{\rm rs, \, LA}$): $\pi(S_{\rm rs}) \sim \mathcal{N}(S_{\rm rs, \, LA}, \sigma_{\rm rs})$. The spectral index $\alpha_{\rm rs}$ was modelled using the empirical distribution determined in \citet{2007mru..confE.140W}: $\pi(\alpha_{\rm rs}) = \mathcal{W}(\alpha_{\rm rs})$.% and \citet{2007yCat..83791442W}: $\pi(\alpha_{\rm rs}) = \mathcal{W}(\alpha_{\rm rs})$.

%%%%%%%%%%%%%%%%%%%%%%%%%%%%%%%%%%%%%%%%%%%%%%%%%%%%%%%%%%%%%%%%%%%%%%%%%%%%%%%

\subsubsection{Likelihood function}

Following \citet{2002MNRAS.334..569H} and FF09, the likelihood function is given by:

\begin{equation}\label{eqn:likelihood} 
\mathcal{L}(\vec{\Theta}) = \frac{1}{Z_{D}}e^{-\frac{1}{2}\chi^{2}}.
\end{equation}
Here $\chi^{2}$ is a measure of the goodness-of-fit between the real and modelled data and can be expressed as
\begin{equation}\label{eqn:chisq}
\chi^{2} = \sum\limits_{\nu,\nu'} (\vec{d}_{\nu} - \vec{d}_{\nu}^{\rm p}(\vec{\Theta}))^{\rm T} \mathbfss{C}_{\nu,\nu'}^{-1} (\vec{d}_{\nu'} - \vec{d}_{\nu'}^{\rm p}(\vec{\Theta})).
\end{equation} 
In this expression $\vec{d}_{\nu}$ are the real data observed by AMI at frequency $\nu$, and $\vec{d}_{\nu}^{\rm p}(\vec{\Theta})$ are the predicted data generated by the model also at frequency $\nu$. AMI data are measured in six neighbouring frequency channels as described in \citet{2008MNRAS.391.1545Z}. 
To generate the predicted data points, values are sampled from $\pi(\vec{\Theta})$ which are used in the calculations outlined in MO12 and \citet{2013MNRAS.430.1344O} to generate a pressure profile for the cluster which can be used to replicate the SZ signal measured by an interferometer as detailed in Sections~4 and~5 of FF09.
$\mathbfss{C}_{\nu,\nu'} \equiv \mathbfss{C}^{\rm ins}_{\nu,\nu'} + \mathbfss{C}^{\rm CMB}_{\nu,\nu'} + \mathbfss{C}^{\rm conf}_{\nu,\nu'}$ is the theoretical covariance matrix, which includes instrumental, primordial CMB and source confusion noise as described in FF09 and \citet{2002MNRAS.334..569H}. Source confusion noise allows for the remaining radio-sources with flux densities less than some flux limit $S_{\rm lim}$, that cannot be measured accurately by the LA. The instrumental noise is actually measured during the observation and so does not need to be predicted. 
Referring back to equation~\ref{eqn:likelihood}, $Z_{D}$ is a normalisation constant given by $(2\pi)^{D / 2} |\mathbfss{C}|^{1/2}$ where $D$ is the length of $\vec{d}$ (the vector of data from all frequencies).

%%%%%%%%%%%%%%%%%%%%%%%%%%%%%%%%%%%%%%%%%%%%%%%%%%%%%%%%%%%%%%%%%%%%%%%%%%%%%%%
%%%%%%%%%%%%%%%%%%%%%%%%%%%%%%%%%%%%%%%%%%%%%%%%%%%%%%%%%%%%%%%%%%%%%%%%%%%%%%%

\subsection{PSZ2 methodology for deriving cluster mass estimates}
\label{subsec:pws} 

For comparison with the mass values obtained with AMI data, we look at the PSZ2 mass estimates obtained from \textit{Planck} data and the requisite scaling relations. The mass values published in PSZ2 are derived from data from one of three detection algorithms: MMF1, MMF3 (%\textcolor{red}{\citealt{2009ApJ...701...32S}; \citealt{2011ApJ...737...61M}} 
both of which are extensions of the matched multi-filter algorithm suitable for SZ studies (MMF, see \citealt{1996MNRAS.279..545H}, \citealt{2002MNRAS.336.1057H} and \citealt{2006A&A...459..341M}), over the whole sky) and PowellSnakes (PwS, \citealt{2012MNRAS.427.1384C}). The former two rely on multi-frequency matched-filter detection methods, whilst PwS is a fully Bayesian method. Since the PwS methodology most closely matches the Bayesian analysis pipeline used for AMI data, we focus on the cluster parameter values from PwS. %Note that in the cases where the only parameter values in PSZ2 were from MMF1 and/ or MMF3, we have to replicate their methodology to derive the PwS estimates, which is described now.

The observable quantity measured by \textit{Planck} is the integrated Comptonisation parameter $Y$. As described in Section~5 of the PSZ2 paper \citep{2016A&A...594A..27P}, for each cluster candidate a two-dimensional posterior for the integrated Comptonisation parameter within the radius $5r_{500}$, $Y(5r_{500})$ and the angular scale radius of the GNFW pressure, $\theta_{\rm p}$ ($= r_{\rm p}/D_{A}$). The values for $Y(5r_{500})$ published in PSZ2 are obtained by marginalising over $\theta_{\rm p}$ and then taking the expected value of $Y(5r_{500})$. We will refer to this value as $Y_{\rm marg}(5r_{500})$.
As described in Sections~5.2 and~5.3 of \citet{2016A&A...594A..27P}, this `blind' measurement of the integrated Comptonisation parameter may not be reliable when the underlying cluster pressure distribution deviates from that given by the GNFW model.
To overcome this, a function relating $Y(5r_{500})$ and $\theta_{\rm p}$ is derived in an attempt to provide prior information on the angular scale of the cluster based on X-ray measurements and earlier \textit{Planck} mission samples. We refer to this function as the slicing function. 

%%%%%%%%%%%%%%%%%%%%%%%%%%%%%%%%%%%%%%%%%%%%%%%%%%%%%%%%%%%%%%%%%%%%%%%%%%%%%%%

\subsubsection{Derivation of the slicing function}

The scaling relations considered here are given in \citet{2014A&A...571A..20P}. Of particular importance to deriving the slicing function, are the $Y(r_{500}) - M(r_{500})$ and $\theta_{500} - M(r_{500})$ relations. The first of these is given by
\begin{equation}\label{eqn:y500m500}
E(z)^{-2/3}\left[\frac{D_{A}^{2}Y(r_{500})}{10^{-4}\rm{Mpc}^{2}}\right] = 10^{-0.19 \pm 0.02} \left[\frac{(1-b)M(r_{500})}{6 \times 10 ^{14}~M_{\rm{Sun}}}\right] ^{1.79 \pm 0.08},
\end{equation}
where $E(z) = \sqrt{\Omega_{\rm M}(1 + z)^{3} + \Omega_{\Lambda}}$ and is equal to the ratio of the Hubble parameter evaluated at redshift $z$ to its value now for a flat $\Lambda$CDM Universe. The factor in the exponent $-2/3$ arises from the scaling relations between mass, temperature and Comptonisation parameter given by equations~1--5 in \citet{2006ApJ...650..128K}. $(1-b)$ represents a bias factor, which is assumed in \citet{2014A&A...571A..20P} to contain four possible observational biases of departure from hydrostatic equilibrium, absolute instrument calibration, temperature inhomogeneities and residual selection bias. Its value is calculated to be $(1 - b) = 0.80 ^{+0.02}_{- 0.01}$ from numerical simulations as described in Appendix~A.4 of \citet{2014A&A...571A..20P}. Equation~\ref{eqn:y500m500} uses the fitting parameters from the relation between $Y_{\rm X}(r_{500})$ (the X-ray `analogue' of the integrated Comptonisation parameter see e.g. \citealt{2006ApJ...650..128K}, $Y_{\rm X}(r_{500}) \equiv M_{\rm g}(r_{500}) T_X$ where $M_{\rm g}$ is the cluster gas mass within $r_{500}$ and $T_X$ is the spectroscopic temperature in the range $[0.15,0.75]r_{500}$) and the X-ray hydrostatic mass, $M_{\rm HE}(r_{500})$ (which is equal to $(1-b) M(r_{500})$), established for 20 local \emph{relaxed} clusters by \citet{2010A&A...517A..92A} to give the relation between the X-ray mass proxy $M_{Y_{X}}(r_{500})$ and $M(r_{500})$. Finally, the fitting parameters for the $Y(r_{500}) - M_{Y_{X}}(r_{500})$ relation are obtained empirically from a 71-cluster sample consisting of SZ data from the \textit{Planck} Early SZ clusters \citep{2011A&A...536A..11P}, of \textit{Planck}-detected LoCuSS clusters \citep{2013A&A...550A.129P} and from the XMM-Newton validation programme \citep{2011A&A...536A...9P}, all with X-ray data taken from XMM-Newton observations (\citealt{2013MNRAS.430..134W} and \citealt{2012MNRAS.423.1024M}).

The $\theta_{500} - M(r_{500})$ relation is based on the equation $M(r_{500}) = 500 \times \frac{4\pi}{3} \rho_{\rm crit}(z) r_{500}^{3}$ and is given by
\begin{equation}\label{eqn:theta500m500}
\theta_{500} = 6.997 \left[\frac{h}{0.7}\right]^{-2/3} \left[ \frac{(1-b)M_{500}}{3 \times 10^{14}~M_{\rm{Sun}}} \right]^{1/3} E(z)^{-2/3}\left[ \frac{D_{A}}{500~\rm{Mpc}} \right].
\end{equation}
Equations~(\ref{eqn:y500m500}) and~(\ref{eqn:theta500m500}) can be solved for $(1-b)M(r_{500})$ and equated to give $Y(r_{500})$ as a function of $\theta_{500}$
\begin{equation}\label{eqn:y500theta500}
Y(r_{500}) = \left[ \frac{\theta_{500}}{6.997} \right] ^{5.4 \pm 0.2} \left[ \frac{h}{0.7} \right] ^{3.60 \pm 0.13} \left[\frac{E(z)^{4.26 \pm 0.13} D_{A}^{3.4 \pm 0.2}}{10^{19.29 \pm 0.54} ~ \rm{Mpc}^{3.4 \pm 0.2}} \right],
\end{equation}
where $Y(r_{500})$ is in $\rm{sr}$. Assuming a GNFW pressure profile, $Y(r_{500})$ can be converted to the corresponding value of $Y(5r_{500})$, through the relation
\begin{equation}\label{eqn:yr500y5r500}
\frac{Y(r_{500})}{Y(5r_{500})} = \frac{B \left( \frac{(c_{500})^{a}}{1+(c_{500})^{a}}; \frac{3 - c}{a}, \frac{b - 3}{a} \right)}{B \left( \frac{(5c_{500})^{a}}{1+(5c_{500})^{a}}; \frac{3 - c}{a}, \frac{b - 3}{a} \right)},
\end{equation} 
where $B(x;y,z) = \int_{0}^{x} t^{y-1}(1-t)^{z-1}\rm{d}t$ is the incomplete beta function. For the GNFW parameter values used in equation~\ref{eqn:gnfw}, equation~\ref{eqn:yr500y5r500} gives a value of $0.55$. Similarly, $\theta_{500}$ can be related to $\theta_{\rm p}$ through the relation $\theta_{\rm p} = \theta_{500} / c_{500} $. 

%%%%%%%%%%%%%%%%%%%%%%%%%%%%%%%%%%%%%%%%%%%%%%%%%%%%%%%%%%%%%%%%%%%%%%%%%%%%%%%

\subsubsection{Mass estimates}
\label{subsubsec:planckmass}
\begin{figure}
  \begin{center}
  \includegraphics[ clip=, width=0.90\linewidth]{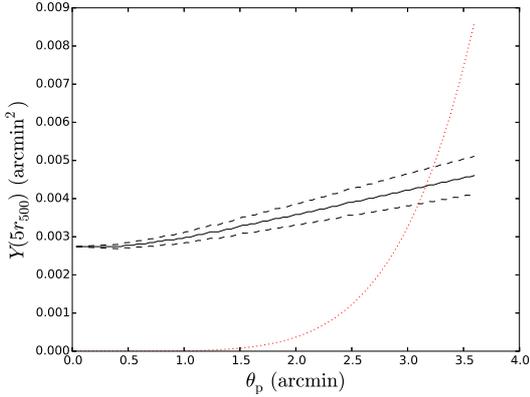}
  \caption{Example of the posterior slicing methodology for cluster PSZ2~G228.16+75.20. The black solid curve represents the `ridge' (i.e. the most probable value of $Y(5r_{500})$ for each $\theta_{\rm p}$) of the posterior. The upper dashed curve represents the upper boundaries of the 68\% maximum likelihood confidence interval on $Y(5r_{500})$ for each value of $\theta_{\rm p}$, and the lower dashed curve corresponds to the lower boundaries. The red dotted curve is the slicing function.}
  \label{graph:slicing}
  \end{center}
\end{figure}

For a given cluster, the resulting $Y(5r_{500})$ function is used to `slice' the posterior, and the value where the function intersects the posterior `ridge' is taken to be the most reliable estimate of $Y(5r_{500})$, given the external information. The posterior ridge (see Figure~\ref{graph:slicing}) is defined to be the value of $Y(5r_{500})$ which gives the highest probability density for a given $\theta_{\rm p}$. The error estimates are obtained by considering where the slicing function intersects with the ridges defined by the 68\% maximum likelihood confidence intervals for $Y(5r_{500})$ at each $\theta_{\rm p}$. $Y(5r_{500})$ is then converted to $Y(r_{500})$ using the the reciprocal of the value given by equation~\ref{eqn:yr500y5r500}, and this is used to derive a value for $M(r_{500})$ using equation~\ref{eqn:y500m500}, but with the $(1-b)$ term excluded. The bias term is not included in the $M(r_{500})$ calculation because it has already been accounted for in the slicing function. Note that this value of $M(r_{500})$ is what is referred to as $M_{\rm SZ}$ in PSZ2.%\citet{2016A&A...594A..27P}.

%%%%%%%%%%%%%%%%%%%%%%%%%%%%%%%%%%%%%%%%%%%%%%%%%%%%%%%%%%%%%%%%%%%%%%%%%%%%%%%
%%%%%%%%%%%%%%%%%%%%%%%%%%%%%%%%%%%%%%%%%%%%%%%%%%%%%%%%%%%%%%%%%%%%%%%%%%%%%%%
%%%%%%%%%%%%%%%%%%%%%%%%%%%%%%%%%%%%%%%%%%%%%%%%%%%%%%%%%%%%%%%%%%%%%%%%%%%%%%%

\section{AMI and PSZ2 mass estimates}
\label{sec:results_i}
First we describe how we arrived at a final sample of clusters for which the AMI mass estimates are compared with those derived from \textit{Planck} data.

%%%%%%%%%%%%%%%%%%%%%%%%%%%%%%%%%%%%%%%%%%%%%%%%%%%%%%%%%%%%%%%%%%%%%%%%%%%%%%%
%%%%%%%%%%%%%%%%%%%%%%%%%%%%%%%%%%%%%%%%%%%%%%%%%%%%%%%%%%%%%%%%%%%%%%%%%%%%%%%

\subsection{Final cluster sample}

%%%%%%%%%%%%%%%%%%%%%%%%%%%%%%%%%%%%%%%%%%%%%%%%%%%%%%%%%%%%%%%%%%%%%%%%%%%%%%%

\subsubsection{Well constrained posterior sample}
\textsc{McAdam} was used on data from the initial sample of 199 clusters. \textsc{MultiNest} failed to produce posterior distributions for two clusters. These clusters were surrounded by high flux, extended radio-sources. Of the 197 clusters for which posterior distributions were produced, 73 clusters show good constraints (adjudged by physical inspection) on the sampling parameters $M(r_{200})$, $f_{\rm gas}(r_{200})$, $x_{\rm c}$ and $y_{\rm c}$; with $z$s ranging from $0.089$ to $0.83$. 
%Similarly, the associated \textit{Planck} S/N ratios spans a wide range relative to the values in Table~\ref{tab:initial_sample}, with minimum and maximum values of $4.97$ and $28.40$ respectively. It is interesting to note the former is only just below the lower limit imposed in YP15. It is also interesting to note that the lowest value of $M_{\rm SZ}$ corresponding to a well constrained cluster is $3.59$, which is almost double the minimum \textit{Planck} mass associated with the initial sample shown in Table~\ref{tab:initial_sample}.

\begin{figure*}
  \begin{center}
  \includegraphics[ width=0.45\linewidth]{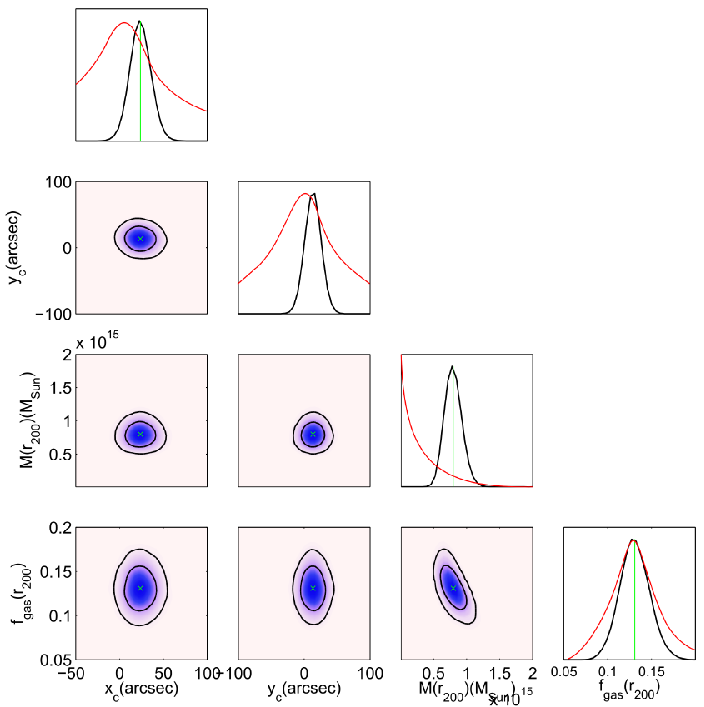} %smaller files for arxiv
  \includegraphics[ width=0.45\linewidth]{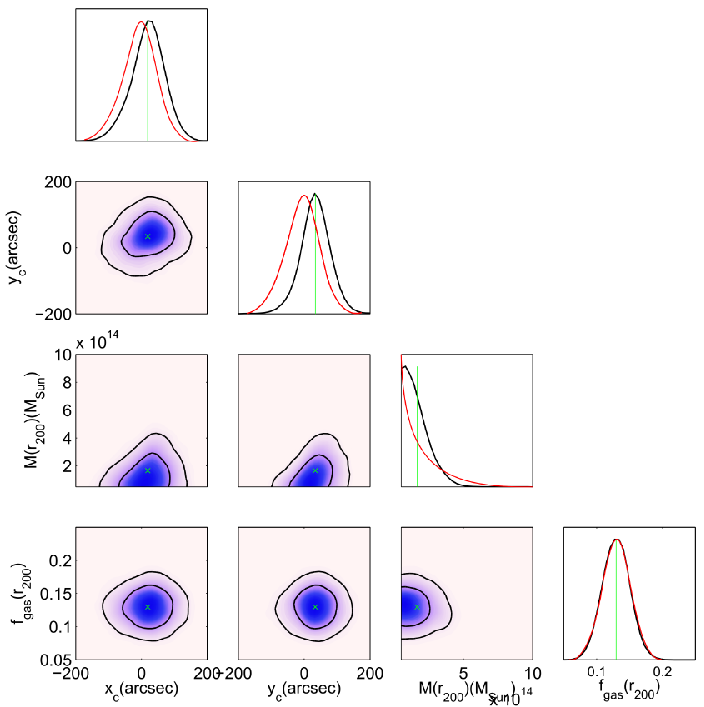}
  \medskip
  \centerline{(a) \hskip 0.45\linewidth (b)}
  \caption{Posterior distributions derived from AMI data for the sampling parameters: $M(r_{200})$; $f_{\rm gas}(r_{200})$; $x_{\rm c}$ and $y_{\rm c}$. The contoured maps show the two-dimensional posteriors for the different pairs of parameters. The contours represent the 95\% and 68\% mean confidence intervals, with the green crosses denoting the expected value of the joint distributions. The four one-dimensional plots are the marginalised posteriors corresponding to the variable given at the bottom of the respective column. The red curves are the prior distributions on the respective parameters. Each green line is the expected value of the distribution. Posterior distributions in (a) show narrow distributions on the cluster mass, with the domain spanning feasible mass values for a galaxy cluster (cluster PSZ2~G184.68+28.91). In such cases the posteriors are said to be well constrained. The mass posteriors in (b) show that the data imply unphysical values for its mass, as the posterior distribution is hitting the lower bound of the prior ($5 \times 10^{13} M_{\rm{Sun}}$) at almost its peak value (cluster PSZ2~G121.77+51.75). The distribution also resembles the uniform in log-space prior assigned to $M(r_{200})$. In such cases the posteriors are said to be poorly constrained with respect to the mass estimates.}
\label{graph:posterior_constraints}
  \end{center}
\end{figure*}

We illustrate a `well constrained' posterior distribution (for cluster PSZ2~G184.68+28.91) in the first half of Figure~\ref{graph:posterior_constraints}, plotted using \textsc{GetDist}\footnote{\url{http://getdist.readthedocs.io/en/latest/}.}. In contrast the second half of Figure~\ref{graph:posterior_constraints} is an example of a cluster (PSZ2~G121.77+51.75) which shows poor constraints on mass as the posterior distribution is peaked at the lower boundary of the mass sampling range ($5\times 10^{13} M_{\rm{Sun}}$) which could not be classed as a detection within our mass prior range. We also note that in the latter case the mass posterior largely resembles the log uniform prior distribution.
%%%%%%%%%%%%%%%%%%%%%%%%%%%%%%%%%%%%%%%%%%%%%%%%%%%%%%%%%%%%%%%%%%%%%%%%%%%%%%%

\subsubsection{Moderate radio-source environment sample}
\begin{figure*}
  \begin{center}
  \includegraphics[ width=0.45\linewidth]{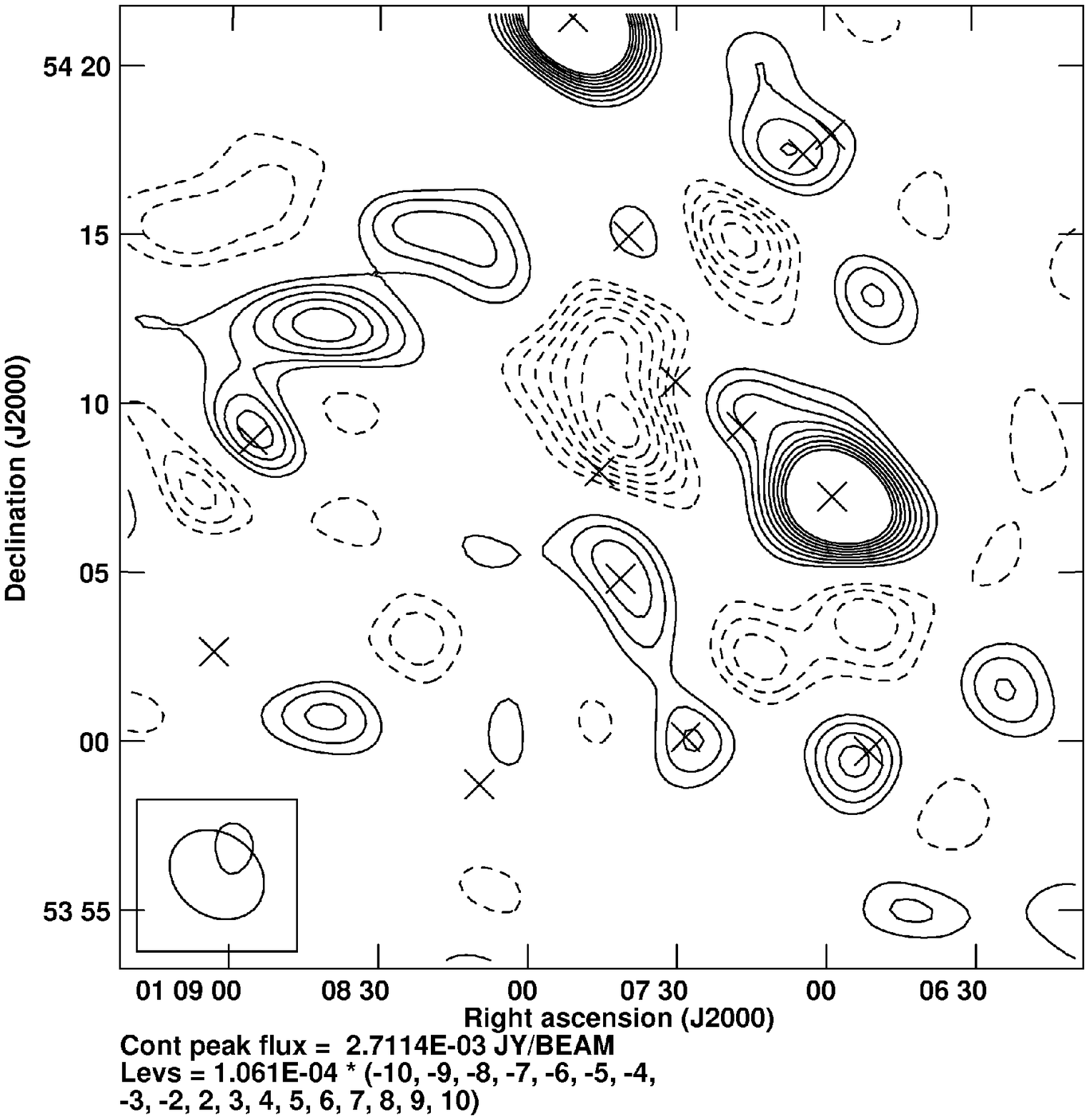}
  \includegraphics[ width=0.45\linewidth]{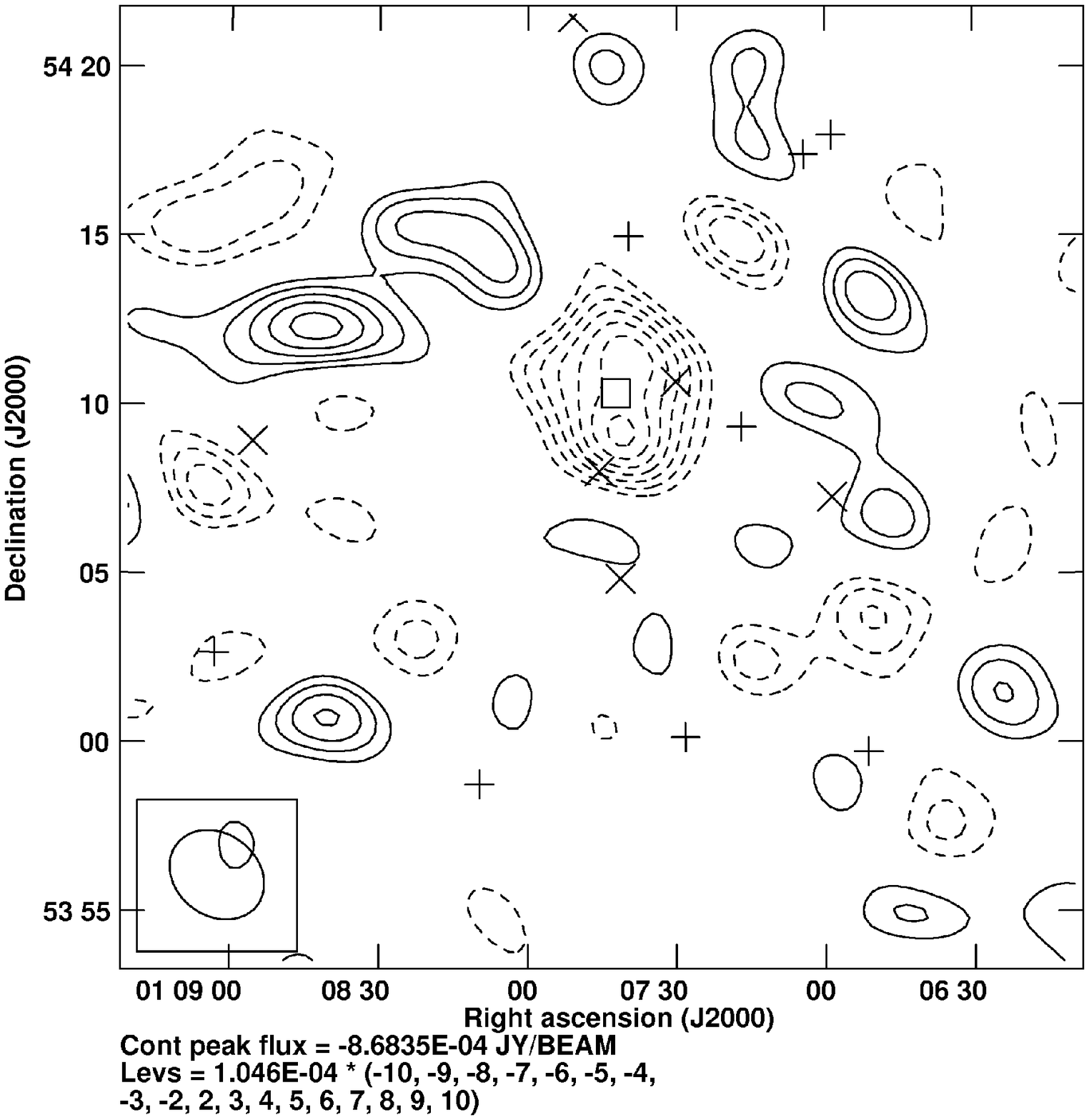}
  \medskip
  \centerline{(a) \hskip 0.45\linewidth (b)}
  \caption{(a) Unsubtracted map produced from AMI observation of cluster PSZ2~G125.37-08.67. Contours are plotted at $\pm (2, 3, 4, ..., 10)\times$ the r.m.s. noise level, and dashed contours are negative. (b) Source subtracted map produced from AMI observation for same cluster. The $\Box$ denotes the \textsc{McAdam}-determined centre of the cluster (posterior mean values for $x_{\rm c}$ and $y_{\rm c}$). Here `$+$' signs denote radio-source positions as measured by the LA which were assigned delta priors on their parameters, whilst `$\times$' denote sources which were assigned priors as described in Section~\ref{subsub:rs_priors}.} 
  \label{graph:bad_rs}
  \end{center}
\end{figure*}

For the 197 cluster sample, AMI data maps were produced using the software package \textsc{AIPS}\footnote{\url{http://aips.nrao.edu/}.} using the automated \textsc{CLEAN} procedure with a limit determined using \textsc{IMEAN}. Source-finding was carried out at a four $\sigma$ level on the LA continuum map, as described in \citet{2011MNRAS.415.1883D} and \citet{2011MNRAS.415.2699A}. For each cluster both a non-source-subtracted and a source-subtracted map was produced. The values used to subtract the sources from the maps were the mean values of the one-dimensional marginalised posterior distributions of the sources' position, flux and spectral index produced by \textsc{McAdam}.
%Two types of maps were produced: unsubtracted and subtracted maps. Unsubtracted maps use data taken by the SA to make a map of the cluster, radio-source, CMB primordial anisotropies and instrumental noise signals. Subtracted maps are the same, but with the radio-source signals `subtracted'. The values which lead to the subtraction are taken from the posterior distributions of the flux and spectral index values of the modelled sources output from \textsc{McAdam}. In other words, the values used are \textsc{McAdam}'s best estimates for the sources' flux and spectral index values, given the prior information measured by the LA, and the fact that all of the surrounding objects have to also be modelled. For more details on how the maps were produced, please refer to \citet{2011arXiv1101.5590A}. \\
Maps of the 73 cluster sample were inspected in detail. It was found that for seven of these clusters, even though the posterior distributions were well constrained, that the radio-source and primordial CMB contamination could bias the cluster parameter estimates in an unpredictable way. In these cases it was found that the subtracted maps contained residual flux close to the cluster centre, from either radio-sources (some of which were extended), radio-frequency interference, or CMB. PSZ2~G125.37-08.67 is an example of one of these clusters and its non-source-subtracted and source-subtracted maps are shown in Figure~\ref{graph:bad_rs}. We thus arrived at a 66 cluster sample.

%%%%%%%%%%%%%%%%%%%%%%%%%%%%%%%%%%%%%%%%%%%%%%%%%%%%%%%%%%%%%%%%%%%%%%%%%%%%%%%

\subsubsection{Well defined cluster-centre sample}
\begin{figure}
  \begin{center}
  \includegraphics[ width=0.90\linewidth]{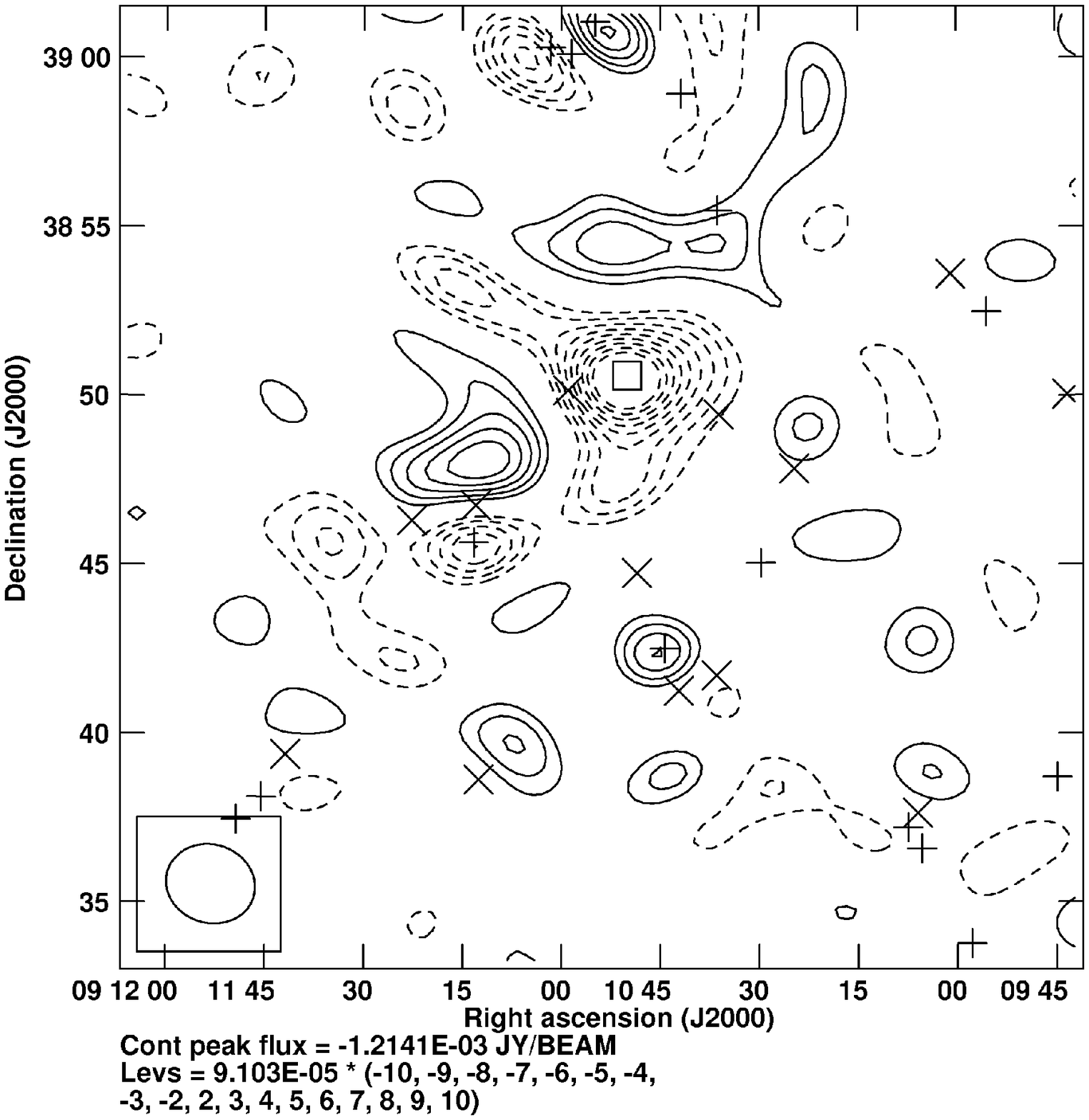}
  \caption{Subtracted map of cluster with ill-defined centre (cluster PSZ2~G183.90+42.99). The cluster is clearly offset from the observation pointing centre (middle of the map), and the lobes to the bottom and the top left of the cluster cause the centre position to be ambiguous.} 
  \label{graph:offset_map}
  \end{center}
\end{figure}

The posteriors of $x_{\rm c}$ and $y_{\rm c}$ give the position of the modelled cluster centre relative to the actual SA pointing centre used for the observation. For seven of the 66 cluster sample, it was found that the mean posterior values of $x_{\rm c}$ and $y_{\rm c}$ changed dramatically between different runs of \textsc{McAdam} (on the same cluster data), by up to $70$ arcseconds in either direction, leading to differences in mass estimates of up to $70\%$. The estimates for these clusters are not reliable, since the model was creating a completely different cluster between runs, and so these clusters were excluded leaving a 59 cluster sample. For the remaining clusters, the change in $M(r_{200})$ between runs was much smaller than the standard deviation of the corresponding posterior distributions. Figure~\ref{graph:offset_map} shows the subtracted map for PSZ2~G183.90+42.99, which we consider to be an example of a cluster with an ill-defined centre. The map shows three flux decrement peaks close to the cluster centre. Movement of the centre between these peaks with the current source environment modelling would lead to a change in the size of the predicted cluster, and consequently different mass estimates each time.

%%%%%%%%%%%%%%%%%%%%%%%%%%%%%%%%%%%%%%%%%%%%%%%%%%%%%%%%%%%%%%%%%%%%%%%%%%%%%%%

\subsubsection{PwS detected cluster sample}
For five of the 59 cluster sample, the data available on the \textit{Planck} website\footnote{\url{https://pla.esac.esa.int/pla/catalogues}.} did not contain a detection using the PwS algorithm, and so no mass estimates based on PwS data could be calculated. Hence the final sample size for which we present the mass estimates from both AMI and \textit{Planck} data is 54. \\

It is important to realise that selection biases are introduced in reducing the sample size from $199$ to $54$. In particular, selecting only the clusters which showed good AMI posterior constraints means that clusters corresponding to a signal too faint for AMI to detect, clusters with large enough angular size for AMI's shortest baselines not to be able to measure the signal from the outskirts of the cluster ("resolved clusters"), and clusters where the radio-source and CMB contamination dwarfs the signal of the cluster, are all likely to have been excluded from the sample to some extent. In addition, removing the seven clusters with an ill defined centre likely removes some unrelaxed clusters from the sample.

%%%%%%%%%%%%%%%%%%%%%%%%%%%%%%%%%%%%%%%%%%%%%%%%%%%%%%%%%%%%%%%%%%%%%%%%%%%%%%%
%%%%%%%%%%%%%%%%%%%%%%%%%%%%%%%%%%%%%%%%%%%%%%%%%%%%%%%%%%%%%%%%%%%%%%%%%%%%%%%

\subsection{AMI and PSZ2 mass estimates}
 
\begin{figure}
  \begin{center}
  \includegraphics[ width=0.90\linewidth]{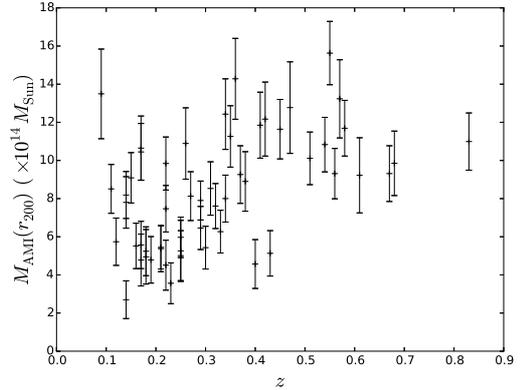}
  \caption{Plot of $M(r_{200})$ derived from AMI data using physical modelling vs redshift for the sample of 54 clusters.}
\label{graph:m200z}
  \end{center}
\end{figure}

The AMI and PSZ2 parameter estimates for the 54 clusters are given in Table~\ref{tab:results}.
The clusters are listed in ascending order of $z$. Note that whether a redshift is photometric or spectroscopic is stated in the fifth column.
All AMI values are the mean values of the corresponding parameter posterior distributions, with the error taken as the standard deviation. The estimates of the sampling parameters are included for comparison with each other, and with the sampling prior ranges and associated parameters given in Table~\ref{tab:clusterpriors}. The AMI values for $M(r_{500})$ are given for comparison with the corresponding PSZ2 estimates.
Two values for the PSZ2 mass estimates are given, $M_{\rm Pl,\, marg}(r_{500})$ and $M_{\rm Pl,\, slice}(r_{500})$. $M_{\rm Pl,\, marg}(r_{500})$ corresponds to the mass given by the $Y(r_{500}) - M(r_{500})$ relation when the marginalised integrated Comptonisation parameter is used as described in Section~\ref{subsec:pws}. The uncertainties associated with these $Y$ values are taken as the standard deviations of the marginalised posteriors. $M_{\rm Pl,\, slice}(r_{500})$ is detailed in Section~\ref{subsubsec:planckmass}; its associated errors are calculated from the $Y(5r_{500})$ values where the slicing function intersects with the two ridges formed by the 68\% confidence interval values of the $Y(5r_{500})$ probability densities over the posterior domain of $\theta_{\rm p}$. \\
Figure~\ref{graph:m200z} shows $M(r_{200})$ as a function of $z$. Excluding the clusters at $z = 0.089,\, 0.4$ and $0.426$, there is a steepening in mass between $ 0.1 \lessapprox z \lessapprox 0.5$ before it flattens off at higher $z$. This result is roughly consistent with the PSZ2 mass estimates (at $r_{500}$) obtained in \citet{2016A&A...594A..27P}.

\begin{figure*}%this is relevant for last Section, but has been put here to stop formatting messing up.
  \begin{center}
  \includegraphics[ width=0.90\linewidth]{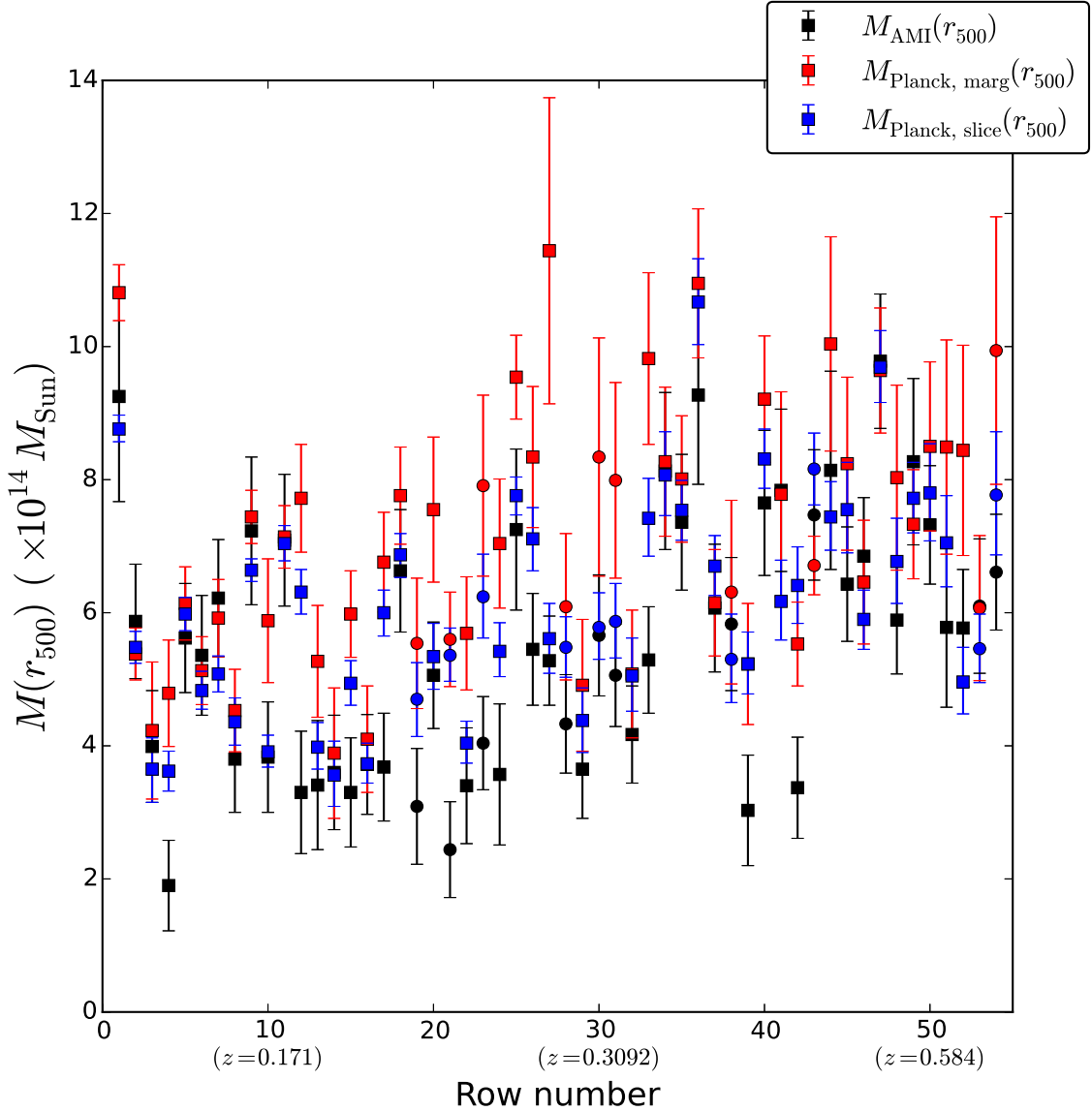}
  \caption{Plot of $M(r_{500})$ vs row number of Table~\ref{tab:results}
 for three different cases: the value derived from AMI data using the physical model, $M_{\rm AMI}(r_{500})$; the value derived from \textit{Planck} data using the marginalised value for $Y(5r_{500})$, $M_{\rm Pl,\, marg}(r_{500})$ and the value derived from \textit{Planck} data using the slicing function value for $Y(5r_{500})$, $M_{\rm Pl,\, slice}(r_{500})$. The row number is monotonically related to $z$, as Table~\ref{tab:results} is sorted by ascending $z$}. The points with circular markers correspond to clusters whose redshifts were measured photometrically (as listed in Table~\ref{tab:results}). 
\label{graph:m500planck}
 \end{center}
\end{figure*}

We now focus on the comparison between AMI and \textit{Planck} mass estimates. Note that \citet{2016A&A...594A..27P} do not provide any means for estimating $M(r_{200})$ from their data, as $r_{200}$ is the distance related to the scale radius ($r_{200} = c_{200} \times r_{\rm s}$) for the NFW dark matter profile given by equation~\ref{eqn:nfw}, which they do not incorporate into their modelling process.
Figure~\ref{graph:m500planck} gives the AMI and two \textit{Planck} estimates for $M(r_{500})$ vs the row number, in Table~\ref{tab:results}. We have not used $z$ as the independent variable in this plot for clarity. The row number is monotonically related to $z$, as Table~\ref{tab:results} is sorted by ascending $z$. 
From Figure~\ref{graph:m500planck} it is clear that AMI underestimates the mass relative to both PSZ2 values. In fact $M(r_{500})$ is lower than $M_{\rm Pl,\, slice}(r_{500})$ in 37 out of 54 cases. $M(r_{500})$ is lower than $M_{\rm Pl,\, marg}(r_{500})$ in 45 out of 54 cases. %This may be due to the physical model prior on $M(r_{200}$ (which is $\propto 1 / M(r_{200}$ in linear-space) which could cause a systematic bias on AMI mass estimates, particularly for low mass clusters.
31 of the AMI masses are within one combined standard deviation of $M_{\rm Pl,\, slice}(r_{500})$, while 46 are within two. Four clusters have discrepancies larger than three combined standard deviations. Three of these clusters are at relatively low redshift ($\leq 0.25$), whilst one is at $z=0.43$.\\

\begin{figure*}%this is relevant for last Section, but has been put here to stop formatting messing up.
  \begin{center}
  \includegraphics[ width=0.90\linewidth]{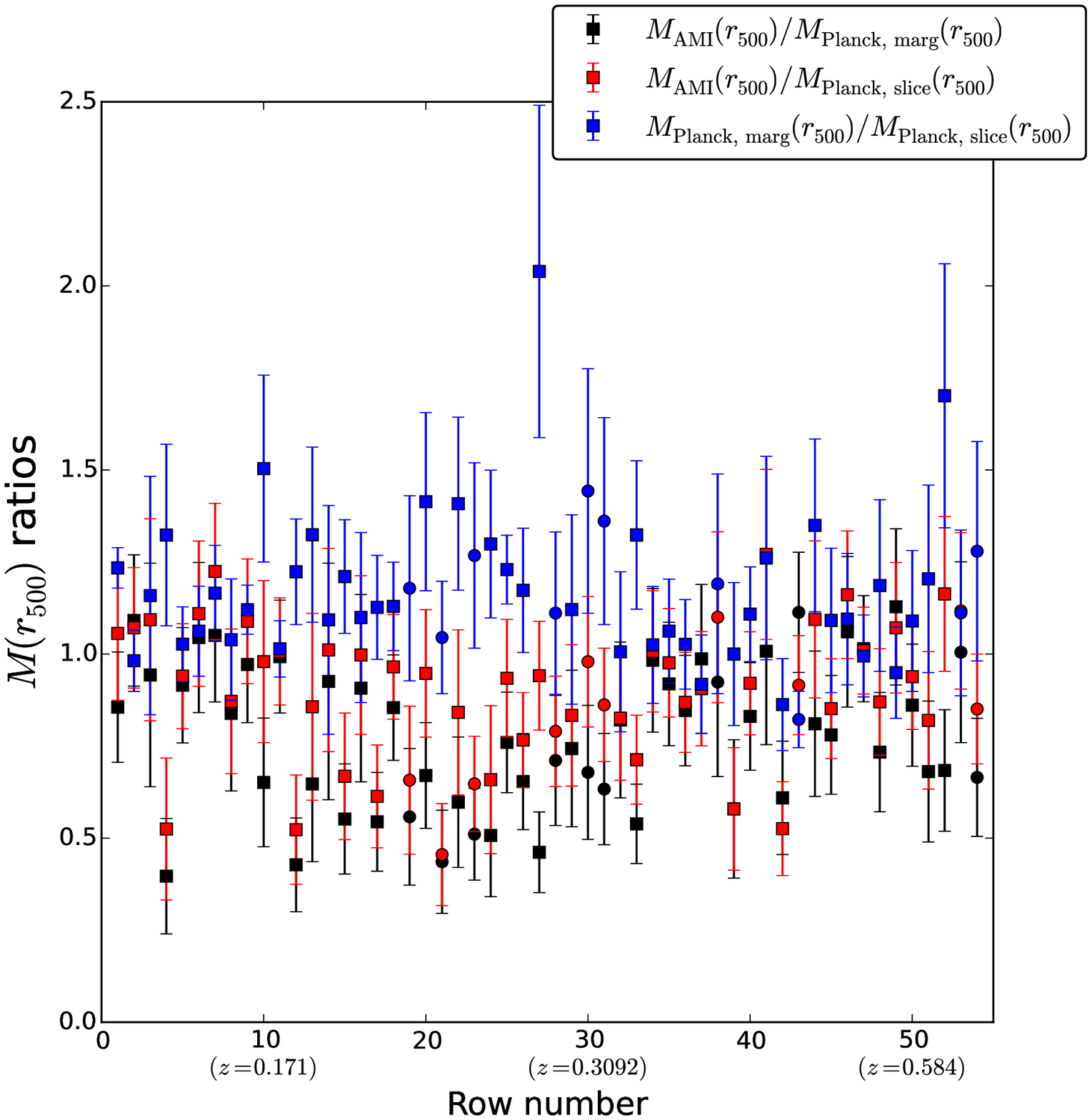}
  \caption{Plot of $M(r_{500})$ ratios vs row number of Table~\ref{tab:results} 
   for three different cases: $M_{\rm AMI}(r_{500}) / M_{\rm Pl,\, marg}(r_{500})$; $M_{\rm AMI}(r_{500}) / M_{\rm Pl,\, slice}(r_{500})$ and $M_{\rm Pl,\, marg}(r_{500}) / M_{\rm Pl,\, slice}(r_{500})$. The points with square markers correspond to clusters whose redshifts were measured spectroscopically, and the circular markers correspond to photometric redshifts (as listed in Table~\ref{tab:results}).
}
\label{graph:m500planckfrac}
  \end{center}
\end{figure*}

Figure~\ref{graph:m500planckfrac} shows the pairwise ratios of mass estimates between the three different methods. The most obvious thing to note is that the ratio of PSZ2 masses is consistently greater than one, which again emphasises the fact that the marginalisation method attributes a much higher mass to the clusters than the slicing method. Furthermore, the ratio of AMI mass to the marginalised mass is small at medium redshift, which suggests that the marginalised mass is systematically high in this range. This graph also emphasises that the AMI mass and the slicing methodology mass are the most consistent with one another.

\newgeometry{margin=1cm} 
\onecolumn
\begin{landscape}

\begin{center}
\begin{longtable}{llllllllllll}
\caption{Summary of values for final sample of 54 clusters. The redshift types correspond to S: spectroscopically measured and P: photometrically measured. $z$, $M(r_{200})$, $x_{\rm c}$, $y_{\rm c}$ and $f_{\rm gas}(r_{200})$ are the physical model sampling parameters. %For clarity, $z$ is given to three decimal places. 
$M_{\rm AMI}(r_{500})$,  $M_{\rm Pl, marg} (r_{500})$ and $M_{\rm Pl, slice} (r_{500})$ are the $M(r_{500})$ estimates obtained from the AMI and \textit{Planck} data respectively. All masses are given in units of $\times 10^{14}~M_{\rm{Sun}}$ and all cluster centre coordinates are measured in arcseconds. }\label{tab:results} \\

\hline \multicolumn{1}{c}{Row} & \multicolumn{1}{c}{\textit{Planck} ID} & \multicolumn{1}{c}{Alias} & \multicolumn{1}{c}{$z$} & \multicolumn{1}{c}{$z$ type} & \multicolumn{1}{c}{$M_{\rm AMI}(r_{200})$} & \multicolumn{1}{c}{$x_{\rm c}$} & \multicolumn{1}{c}{$y_{\rm c}$} & \multicolumn{1}{c}{$f_{\rm gas}(r_{200})$} & \multicolumn{1}{c}{$M_{\rm AMI}(r_{500})$} & \multicolumn{1}{c}{$M_{\rm Pl, marg}(r_{500})$} & \multicolumn{1}{c}{$M_{\rm Pl, slice} (r_{500})$} \\ \hline 
\endfirsthead

\multicolumn{12}{c}%
{{\tablename\ \thetable{} -- continued from previous page}} \\
\hline \multicolumn{1}{c}{Row} & \multicolumn{1}{c}{\textit{Planck} ID} & \multicolumn{1}{c}{Alias} & \multicolumn{1}{c}{$z$} & \multicolumn{1}{c}{$z$ type} & \multicolumn{1}{c}{$M_{\rm AMI}(r_{200})$} & \multicolumn{1}{c}{$x_{\rm c}$} & \multicolumn{1}{c}{$y_{\rm c}$} & \multicolumn{1}{c}{$f_{\rm gas}(r_{200})$} & \multicolumn{1}{c}{$M_{\rm AMI}(r_{500})$} & \multicolumn{1}{c}{$M_{\rm Pl, marg}(r_{500})$} & \multicolumn{1}{c}{$M_{\rm Pl, slice} (r_{500})$} \\ \hline 
\endhead

\tabulinesep=_1mm
\extrarowsep=1mm
\LTcapwidth=\textwidth

1 & PSZ2~G044.20+48.66 & ACO2142 & $ 0.0894  $ & S   & $ 13.49 \pm 2.35  $ & $ 9.14  \pm 18.20 $ & $ 8.80  \pm 15.08 $ & $ 0.13  \pm 0.02  $ & $ 9.25  \pm 1.58  $ & $ 10.81 \pm 0.42  $ & $ 8.76  \pm ^{  0.19  } _{  0.21  } $ \\
2 & PSZ2~G053.53+59.52 & ACO2034 & $ 0.113 $ & S   & $ 8.51  \pm 1.28  $ & $ -1.80 \pm 13.10 $ & $ 19.39 \pm 9.86  $ & $ 0.13  \pm 0.02  $ & $ 5.87  \pm 0.86  $ & $ 5.38  \pm 0.39  $ & $ 5.48  \pm ^{  0.24  } _{  0.24  } $ \\
3 & PSZ2~G151.90+11.63 & CIZAJ0515.3+5845  & $ 0.12  $ & S   & $ 5.74  \pm 1.24  $ & $ 67.58 \pm 27.09 $ & $ 68.01 \pm 18.58 $ & $ 0.13  \pm 0.02  $ & $ 3.99  \pm 0.84  $ & $ 4.23  \pm 1.03  $ & $ 3.65  \pm ^{  0.50  } _{  0.47  } $ \\
4 & PSZ2~G218.59+71.31 & ACO1272 & $ 0.137 $ & S   & $ 2.70  \pm 0.99  $ & $ 2.82  \pm 25.21 $ & $ -16.62  \pm 25.98 $ & $ 0.13  \pm 0.02  $ & $ 1.90  \pm 0.68  $ & $ 4.79  \pm 0.80  $ & $ 3.62  \pm ^{  0.30  } _{  0.30  } $ \\
5 & PSZ2~G226.18+76.79 & ACO1413 & $ 0.1427  $ & S   & $ 8.19  \pm 1.23  $ & $ -35.33  \pm 10.98 $ & $ -1.13 \pm 13.44 $ & $ 0.13  \pm 0.02  $ & $ 5.62  \pm 0.82  $ & $ 6.14  \pm 0.55  $ & $ 5.98  \pm ^{  0.25  } _{  0.25  } $ \\
6 & PSZ2~G165.06+54.13 & ACO990  & $ 0.144 $ & S   & $ 7.80  \pm 1.35  $ & $ 32.43 \pm 13.21 $ & $ -27.57  \pm 15.52 $ & $ 0.14  \pm 0.02  $ & $ 5.36  \pm 0.90  $ & $ 5.13  \pm 0.51  $ & $ 4.83  \pm ^{  0.28  } _{  0.29  } $ \\
7 & PSZ2~G077.90-26.63 & ACO2409   & $ 0.147 $ & S   & $ 9.09  \pm 1.32  $ & $ -26.87  \pm 10.89 $ & $ 18.00 \pm 11.85 $ & $ 0.14  \pm 0.02  $ & $ 6.22  \pm 0.88  $ & $ 5.92  \pm 0.58  $ & $ 5.08  \pm ^{  0.27  } _{  0.27  } $ \\
8 & PSZ2~G050.40+31.17 & ACO2259 & $ 0.164 $ & S   & $ 5.52  \pm 1.19  $ & $ 35.72 \pm 21.77 $ & $ 9.31  \pm 19.56 $ & $ 0.13  \pm 0.02  $ & $ 3.80  \pm 0.80  $ & $ 4.53  \pm 0.62  $ & $ 4.36  \pm ^{  0.35  } _{  0.36  } $ \\
9 & PSZ2~G097.72+38.12 & ACO2218   & $ 0.1709  $ & S   & $ 10.65 \pm 1.68  $ & $ 31.99 \pm 15.25 $ & $ -0.95 \pm 13.52 $ & $ 0.13  \pm 0.02  $ & $ 7.23  \pm 1.11  $ & $ 7.44  \pm 0.40  $ & $ 6.64  \pm ^{  0.17  } _{  0.17  } $ \\
10  & PSZ2~G099.30+20.92 & MCXCJ1935.3+6734  & $ 0.171 $ & S   & $ 5.57  \pm 1.24  $ & $ -37.19  \pm 19.92 $ & $ -24.50  \pm 21.16 $ & $ 0.13  \pm 0.02  $ & $ 3.83  \pm 0.83  $ & $ 5.88  \pm 0.93  $ & $ 3.91  \pm ^{  0.23  } _{  0.25  } $ \\
11  & PSZ2~G067.17+67.46 & ACO1914 & $ 0.1712  $ & S   & $ 10.45 \pm 1.49  $ & $ 31.39 \pm 12.81 $ & $ -33.15  \pm 11.99 $ & $ 0.13  \pm 0.02  $ & $ 7.09  \pm 0.99  $ & $ 7.14  \pm 0.47  $ & $ 7.04  \pm ^{  0.26  } _{  0.27  } $ \\
12  & PSZ2~G167.67+17.63 & RXJ0638.1+4747  & $ 0.174 $ & S   & $ 4.78  \pm 1.36  $ & $ -28.70  \pm 31.24 $ & $ 10.76 \pm 28.64 $ & $ 0.13  \pm 0.02  $ & $ 3.30  \pm 0.92  $ & $ 7.72  \pm 0.81  $ & $ 6.31  \pm ^{  0.33  } _{  0.34  } $ \\
13  & PSZ2~G066.68+68.44 & ACO1902 & $ 0.181 $ & S   & $ 4.95  \pm 1.43  $ & $ 56.07 \pm 25.47 $ & $ 8.14  \pm 33.23 $ & $ 0.13  \pm 0.02  $ & $ 3.41  \pm 0.97  $ & $ 5.27  \pm 0.84  $ & $ 3.98  \pm ^{  0.33  } _{  0.37  } $ \\
14  & PSZ2~G065.28+44.53 & ACO2187 & $ 0.183 $ & S   & $ 5.24  \pm 1.28  $ & $ -16.66  \pm 22.61 $ & $ -16.54  \pm 21.65 $ & $ 0.13  \pm 0.02  $ & $ 3.60  \pm 0.86  $ & $ 3.89  \pm 0.98  $ & $ 3.56  \pm ^{  0.47  } _{  0.51  } $ \\
15  & PSZ2~G084.47+12.63 & MCXCJ1948.3+5113  & $ 0.185 $ & S   & $ 4.79  \pm 1.22  $ & $ -73.73  \pm 31.17 $ & $ -16.97  \pm 20.93 $ & $ 0.13  \pm 0.02  $ & $ 3.30  \pm 0.82  $ & $ 5.98  \pm 0.65  $ & $ 4.94  \pm ^{  0.33  } _{  0.34  } $ \\
16  & PSZ2~G100.04+23.73 & ACO2317   & $ 0.21  $ & S   & $ 5.44  \pm 1.13  $ & $ 20.24 \pm 19.02 $ & $ -22.73  \pm 20.90 $ & $ 0.13  \pm 0.02  $ & $ 3.72  \pm 0.75  $ & $ 4.10  \pm 0.80  $ & $ 3.73  \pm ^{  0.29  } _{  0.31  } $ \\
17  & PSZ2~G180.60+76.65 & SDSSCGB26344.3  & $ 0.2138  $ & S   & $ 5.38  \pm 1.21  $ & $ 37.81 \pm 15.59 $ & $ -66.98  \pm 19.41 $ & $ 0.13  \pm 0.02  $ & $ 3.68  \pm 0.81  $ & $ 6.76  \pm 0.75  $ & $ 6.00  \pm ^{  0.35  } _{  0.34  } $ \\
18  & PSZ2~G166.09+43.38 & ACO773N & $ 0.2172  $ & S   & $ 9.84  \pm 1.39  $ & $ -5.35 \pm 10.66 $ & $ -3.98 \pm 9.70  $ & $ 0.13  \pm 0.02  $ & $ 6.63  \pm 0.92  $ & $ 7.76  \pm 0.73  $ & $ 6.87  \pm ^{  0.34  } _{  0.32  } $ \\
19  & PSZ2~G125.30-27.99 & N/A & $ 0.223 $ & P & $ 4.51  \pm 1.31  $ & $ -8.08 \pm 26.99 $ & $ 8.82  \pm 30.24 $ & $ 0.13  \pm 0.02  $ & $ 3.09  \pm 0.87  $ & $ 5.54  \pm 0.98  $ & $ 4.70  \pm ^{  0.56  } _{  0.55  } $ \\
20  & PSZ2~G060.13+11.44 & N/A & $ 0.224 $ & S   & $ 7.47  \pm 1.22  $ & $ -64.79  \pm 12.50 $ & $ -49.27  \pm 14.16 $ & $ 0.13  \pm 0.02  $ & $ 5.06  \pm 0.80  $ & $ 7.55  \pm 1.09  $ & $ 5.34  \pm ^{  0.49  } _{  0.50  } $ \\
21  & PSZ2~G166.62+42.13 & ACO746  & $ 0.232 $ & P   & $ 3.56  \pm 1.07  $ & $ -38.98  \pm 29.87 $ & $ -38.09  \pm 37.84 $ & $ 0.13  \pm 0.02  $ & $ 2.44  \pm 0.72  $ & $ 5.60  \pm 0.71  $ & $ 5.36  \pm ^{  0.39  } _{  0.41  } $ \\
22  & PSZ2~G097.94+19.43 & 4C 65.28  & $ 0.25  $ & S   & $ 5.01  \pm 1.31  $ & $ -114.76 \pm 22.50 $ & $ -13.64  \pm 34.07 $ & $ 0.13  \pm 0.02  $ & $ 3.40  \pm 0.87  $ & $ 5.69  \pm 0.85  $ & $ 4.04  \pm ^{  0.30  } _{  0.33  } $ \\
23  & PSZ2~G164.29+08.94 & N/A & $ 0.251 $ & P   & $ 5.97  \pm 1.06  $ & $ -62.17  \pm 14.03 $ & $ 18.12 \pm 17.06 $ & $ 0.13  \pm 0.02  $ & $ 4.04  \pm 0.70  $ & $ 7.91  \pm 1.36  $ & $ 6.24  \pm ^{  0.62  } _{  0.64  } $ \\
24  & PSZ2~G133.60+69.04 & RXJ1229.0+4737  & $ 0.254 $ & S   & $ 5.26  \pm 1.60  $ & $ 5.87  \pm 25.04 $ & $ 59.40 \pm 37.35 $ & $ 0.13  \pm 0.02  $ & $ 3.57  \pm 1.06  $ & $ 7.04  \pm 0.97  $ & $ 5.42  \pm ^{  0.38  } _{  0.43  } $ \\
25  & PSZ2~G086.47+15.31 & MCXCJ1938.3+5409  & $ 0.26  $ & S   & $ 10.89 \pm 1.87  $ & $ -39.65  \pm 13.24 $ & $ 19.83 \pm 12.61 $ & $ 0.13  \pm 0.02  $ & $ 7.25  \pm 1.21  $ & $ 9.54  \pm 0.63  $ & $ 7.76  \pm ^{  0.29  } _{  0.28  } $ \\
26  & PSZ2~G139.62+24.18 & N/A & $ 0.2671  $ & S   & $ 8.13  \pm 1.28  $ & $ 36.66 \pm 11.64 $ & $ -12.58  \pm 10.80 $ & $ 0.13  \pm 0.02  $ & $ 5.45  \pm 0.84  $ & $ 8.34  \pm 1.06  $ & $ 7.11  \pm ^{  0.48  } _{  0.47  } $ \\
27  & PSZ2~G184.68+28.91 & ACO611  & $ 0.288 $ & S   & $ 7.90  \pm 1.02  $ & $ 22.61 \pm 10.45 $ & $ 13.48 \pm 9.97  $ & $ 0.13  \pm 0.02  $ & $ 5.28  \pm 0.67  $ & $ 11.44 \pm 2.30  $ & $ 5.61  \pm ^{  0.52  } _{  0.53  } $ \\
28  & PSZ2~G154.13+40.19 & ACO747  & $ 0.29  $ & P & $ 6.46  \pm 1.13  $ & $ 70.99 \pm 14.72 $ & $ -42.86  \pm 13.25 $ & $ 0.13  \pm 0.02  $ & $ 4.33  \pm 0.74  $ & $ 6.09  \pm 1.10  $ & $ 5.48  \pm ^{  0.45  } _{  0.46  } $ \\
29  & PSZ2~G095.49+16.41 & N/A & $ 0.3 $ & S   & $ 5.43  \pm 1.12  $ & $ -24.47  \pm 19.10 $ & $ -102.18 \pm 18.33 $ & $ 0.13  \pm 0.02  $ & $ 3.65  \pm 0.74  $ & $ 4.91  \pm 0.99  $ & $ 4.38  \pm ^{  0.48  } _{  0.49  } $ \\
30  & PSZ2~G109.52-19.16 & N/A & $ 0.3092  $ & P   & $ 8.53  \pm 1.40  $ & $ -30.38  \pm 13.77 $ & $ -15.21  \pm 15.15 $ & $ 0.13  \pm 0.02  $ & $ 5.66  \pm 0.91  $ & $ 8.34  \pm 1.79  $ & $ 5.78  \pm ^{  0.48  } _{  0.52  } $ \\
31  & PSZ2~G198.90+18.16 & [SPD2011] 298 & $ 0.3184  $ & P   & $ 7.61  \pm 1.18  $ & $ 26.76 \pm 14.62 $ & $ -58.07  \pm 11.95 $ & $ 0.13  \pm 0.02  $ & $ 5.06  \pm 0.77  $ & $ 7.99  \pm 1.47  $ & $ 5.87  \pm ^{  0.55  } _{  0.57  } $ \\
32  & PSZ2~G152.33+81.28 & MCXCJ1230.7+3439  & $ 0.333 $ & S   & $ 6.27  \pm 1.12  $ & $ -52.81  \pm 20.78 $ & $ 44.11 \pm 14.62 $ & $ 0.13  \pm 0.02  $ & $ 4.17  \pm 0.73  $ & $ 5.08  \pm 0.96  $ & $ 5.05  \pm ^{  0.53  } _{  0.57  } $ \\
33  & PSZ2~G108.17-11.56 & N/A & $ 0.336 $ & S   & $ 8.00  \pm 1.23  $ & $ 35.19 \pm 13.14 $ & $ -70.15  \pm 19.09 $ & $ 0.13  \pm 0.02  $ & $ 5.29  \pm 0.80  $ & $ 9.82  \pm 1.29  $ & $ 7.42  \pm ^{  0.57  } _{  0.60  } $ \\
34  & PSZ2~G132.47-17.27 & MCXCJ0142.9+4438  & $ 0.341 $ & S   & $ 12.43 \pm 1.85  $ & $ 31.87 \pm 10.19 $ & $ 15.27 \pm 12.93 $ & $ 0.13  \pm 0.02  $ & $ 8.13  \pm 1.18  $ & $ 8.27  \pm 1.12  $ & $ 8.07  \pm ^{  0.61  } _{  0.65  } $ \\
35  & PSZ2~G207.88+81.31 & ACO1489 & $ 0.353 $ & S   & $ 11.26 \pm 1.61  $ & $ 68.55 \pm 8.44  $ & $ 62.56 \pm 11.55 $ & $ 0.13  \pm 0.02  $ & $ 7.36  \pm 1.02  $ & $ 8.01  \pm 0.95  $ & $ 7.54  \pm ^{  0.45  } _{  0.45  } $ \\
36  & PSZ2~G157.32-26.77 & MCSJ0308.9+2645   & $ 0.356 $ & S   & $ 14.28 \pm 2.12  $ & $ 0.33  \pm 8.12  $ & $ 17.65 \pm 11.53 $ & $ 0.13  \pm 0.02  $ & $ 9.27  \pm 1.34  $ & $ 10.95 \pm 1.12  $ & $ 10.67 \pm ^{  0.64  } _{  0.65  } $ \\
37  & PSZ2~G071.21+28.86 & RXSJ175201.5+444046   & $ 0.366 $ & S   & $ 9.26  \pm 1.51  $ & $ -29.82  \pm 9.95  $ & $ -12.58  \pm 13.26 $ & $ 0.13  \pm 0.02  $ & $ 6.07  \pm 0.96  $ & $ 6.15  \pm 0.80  $ & $ 6.70  \pm ^{  0.44  } _{  0.46  } $ \\
38  & PSZ2~G194.98+54.12 & MCSJ1006.9+3200 & $ 0.375 $ & P & $ 8.90  \pm 1.56  $ & $ 32.58 \pm 12.17 $ & $ -0.22 \pm 19.18 $ & $ 0.13  \pm 0.02  $ & $ 5.83  \pm 1.00  $ & $ 6.31  \pm 1.38  $ & $ 5.30  \pm ^{  0.65  } _{  0.68  } $ \\
39  & PSZ2~G109.86+27.94 & N/A & $ 0.4 $ & S   & $ 4.57  \pm 1.28  $ & $ 3.98  \pm 22.50 $ & $ 7.39  \pm 18.70 $ & $ 0.13  \pm 0.02  $ & $ 3.03  \pm 0.83  $ & $ 5.23  \pm 0.91  $ & $ 5.23  \pm ^{  0.45  } _{  0.48  } $ \\
40  & PSZ2~G083.29-31.03 & MCXCJ2228.6+2036  & $ 0.412 $ & S   & $ 11.85 \pm 1.73  $ & $ 81.05 \pm 13.29 $ & $ -3.42 \pm 12.73 $ & $ 0.13  \pm 0.02  $ & $ 7.65  \pm 1.09  $ & $ 9.21  \pm 0.95  $ & $ 8.31  \pm ^{  0.44  } _{  0.45  } $ \\
41  & PSZ2~G063.38+53.44 & NSCJ1537+392702   & $ 0.422 $ & S   & $ 12.17 \pm 1.94  $ & $ 46.13 \pm 12.01 $ & $ 46.02 \pm 9.37  $ & $ 0.13  \pm 0.02  $ & $ 7.84  \pm 1.22  $ & $ 7.78  \pm 1.54  $ & $ 6.17  \pm ^{  0.58  } _{  0.62  } $ \\
42  & PSZ2~G063.80+11.42 & N/A & $ 0.426 $ & S   & $ 5.13  \pm 1.19  $ & $ -36.41  \pm 22.22 $ & $ -47.14  \pm 19.79 $ & $ 0.13  \pm 0.02  $ & $ 3.37  \pm 0.76  $ & $ 5.53  \pm 0.63  $ & $ 6.41  \pm ^{  0.57  } _{  0.58  } $ \\
43  & PSZ2~G157.43+30.34 & RXJ0748.6+5940  & $ 0.45  $ & P   & $ 11.64 \pm 1.56  $ & $ -61.32  \pm 7.38  $ & $ 4.53  \pm 8.27  $ & $ 0.13  \pm 0.02  $ & $ 7.47  \pm 0.98  $ & $ 6.71  \pm 0.44  $ & $ 8.16  \pm ^{  0.54  } _{  0.54  } $ \\
44  & PSZ2~G150.56+58.32 & CLGJ1115+5319 & $ 0.47  $ & S   & $ 12.77 \pm 2.40  $ & $ 10.18 \pm 13.31 $ & $ 34.06 \pm 18.57 $ & $ 0.13  \pm 0.02  $ & $ 8.14  \pm 1.49  $ & $ 10.04 \pm 1.61  $ & $ 7.44  \pm ^{  0.50  } _{  0.53  } $ \\
45  & PSZ2~G170.98+39.45 & [SPD2011] 16774 & $ 0.5131  $ & S   & $ 10.11 \pm 1.38  $ & $ 31.48 \pm 10.20 $ & $ -30.87  \pm 12.67 $ & $ 0.12  \pm 0.02  $ & $ 6.43  \pm 0.86  $ & $ 8.24  \pm 1.30  $ & $ 7.55  \pm ^{  0.65  } _{  0.71  } $ \\
46  & PSZ2~G094.56+51.03 & N/A & $ 0.5392  $ & S   & $ 10.83 \pm 1.43  $ & $ 81.61 \pm 8.09  $ & $ 52.86 \pm 8.80  $ & $ 0.13  \pm 0.02  $ & $ 6.85  \pm 0.88  $ & $ 6.46  \pm 0.93  $ & $ 5.90  \pm ^{  0.45  } _{  0.44  } $ \\
47  & PSZ2~G228.16+75.20 & CLGJ1149+2223   & $ 0.545 $ & S   & $ 15.63 \pm 1.66  $ & $ -15.49  \pm 5.32  $ & $ 17.11 \pm 4.75  $ & $ 0.13  \pm 0.01  $ & $ 9.78  \pm 1.01  $ & $ 9.64  \pm 0.94  $ & $ 9.69  \pm ^{  0.53  } _{  0.55  } $ \\
48  & PSZ2~G213.39+80.59 & SDSSCGB41791  & $ 0.5586  $ & S   & $ 9.31  \pm 1.32  $ & $ -9.73 \pm 11.90 $ & $ 69.37 \pm 12.14 $ & $ 0.13  \pm 0.02  $ & $ 5.89  \pm 0.81  $ & $ 8.03  \pm 1.39  $ & $ 6.77  \pm ^{  0.63  } _{  0.65  } $ \\
49  & PSZ2~G066.41+27.03 & N/A & $ 0.5699  $ & S   & $ 13.23 \pm 2.05  $ & $ -33.18  \pm 11.12 $ & $ 97.03 \pm 11.32 $ & $ 0.13  \pm 0.02  $ & $ 8.27  \pm 1.25  $ & $ 7.33  \pm 0.82  $ & $ 7.72  \pm ^{  0.52  } _{  0.54  } $ \\
50  & PSZ2~G144.83+25.11 & CLGJ0647+7015   & $ 0.584 $ & S   & $ 11.69 \pm 1.46  $ & $ 4.15  \pm 7.87  $ & $ -1.21 \pm 8.54  $ & $ 0.13  \pm 0.02  $ & $ 7.32  \pm 0.89  $ & $ 8.50  \pm 1.27  $ & $ 7.80  \pm ^{  0.72  } _{  0.74  } $ \\
51  & PSZ2~G045.87+57.70 & N/A & $ 0.611 $ & S   & $ 9.22  \pm 1.97  $ & $ 11.71 \pm 14.87 $ & $ 24.21 \pm 12.21 $ & $ 0.13  \pm 0.02  $ & $ 5.78  \pm 1.20  $ & $ 8.49  \pm 1.61  $ & $ 7.05  \pm ^{  0.66  } _{  0.71  } $ \\
52  & PSZ2~G108.27+48.66 & N/A & $ 0.674 $ & S   & $ 9.31  \pm 1.46  $ & $ 9.99  \pm 11.34 $ & $ 35.79 \pm 11.45 $ & $ 0.13  \pm 0.02  $ & $ 5.77  \pm 0.88  $ & $ 8.44  \pm 1.58  $ & $ 4.96  \pm ^{  0.48  } _{  0.52  } $ \\
53  & PSZ2~G086.93+53.18 & N/A & $ 0.6752  $ & P   & $ 9.85  \pm 1.69  $ & $ -47.72  \pm 14.38 $ & $ 27.69 \pm 10.67 $ & $ 0.13  \pm 0.02  $ & $ 6.10  \pm 1.01  $ & $ 6.07  \pm 1.09  $ & $ 5.46  \pm ^{  0.51  } _{  0.52  } $ \\
54  & PSZ2~G141.77+14.19 & N/A & $ 0.83  $ & P   & $ 10.99 \pm 1.50  $ & $ -4.36 \pm 8.54  $ & $ -19.02  \pm 8.85  $ & $ 0.13  \pm 0.02  $ & $ 6.61  \pm 0.87  $ & $ 9.94  \pm 2.01  $ & $ 7.77  \pm ^{  0.90  } _{  0.95  } $ \\

\hline

\end{longtable}
\end{center}

\end{landscape}
\restoregeometry

%%%%%%%%%%%%%%%%%%%%%%%%%%%%%%%%%%%%%%%%%%%%%%%%%%%%%%%%%%%%%%%%%%%%%%%%%%%%%%%
%%%%%%%%%%%%%%%%%%%%%%%%%%%%%%%%%%%%%%%%%%%%%%%%%%%%%%%%%%%%%%%%%%%%%%%%%%%%%%%
%%%%%%%%%%%%%%%%%%%%%%%%%%%%%%%%%%%%%%%%%%%%%%%%%%%%%%%%%%%%%%%%%%%%%%%%%%%%%%%

\section{AMI simulations with PSZ2 mass inputs}
\label{sec:results_ii}
%errors (noise contributions) aren't necessarily gaussian, and we are not considering the sum of random variables in these histograms, we are simply considering draws from different cluster simulations. Thus we cannot expect the histograms to be gaussian. If it were gaussian (the errors were all gaussian), then even though the clusters of simulations are different (not identically distributed), the fact that in each case we are subtracting the true mean, would make each cluster simulation to be gaussian distributed with zero mean, and since we are dividing by the standard deviations, unit variance. Thus each cluster simulation would be gaussian distributed with zero mean, and hence even though the simulations are different, with this transformation they would be iid gaussian distributed with zero mean. However, this would not be the case for the real source environment, since we are adding different biases to each simulation, and thus they wouldn't be iid, and hence wouldn't be gauss distributed with zero mean and unit variance (each simulation has a different mean and a different standard deviation. whether they are gaussian, would depend on rest of errors again).
To investigate further the discrepancies between the mass estimates, it was decided to create simulated data based on the PSZ2 mass estimates obtained from the slicing methodology, which were then `observed' by AMI. The data from these simulated observations were analysed the same way as the real data.
The simulations were carried out using the in-house AMI simulation package \textsc{Profile}, which has been used in various forms in e.g. \citet{2002MNRAS.333..318G}, \citet{2011MNRAS.415.2708A}, \citet{2012MNRAS.421.1136A} and \citet{2013MNRAS.430.1344O}.
The input parameters for the simulation-- which uses the physical model to create the cluster-- are the sampling parameters of the model. Since \citet{2016A&A...594A..27P} does not give a method for calculating $M(r_{200})$ it was calculated as follows. First $r_{500}$ was calculated by solving $M_{\rm SZ} = 500 \times \frac{4\pi}{3} \rho_{\rm crit}(z) r_{500}^{3}$ for $r_{500}$. $r_{200}$ can be determined from $r_{500}$, but we note that the function mapping from $r_{200}$ to $r_{500}$ is non-invertible, thus $r_{200}$ had to be calculated by solving equation~\ref{eqn:r200r5001} iteratively. $M(r_{200})$ can then be calculated from $M(r_{200}) = 200 \times \frac{4\pi}{3} \rho_{\rm crit}(z) r_{200}^{3}$. \\
As well as the values of $M(r_{200})$ derived from PSZ2 mass estimates, values for the other inputs were also required. We used $f_{\rm gas}(r_{200}) = 0.13$, $z = z_{\rm Planck}$, and $x_{\rm c} = y_{\rm c} = 0$~arcsec.\\%, so that the replicated radio-source environment (see Section~\ref{subsec:sourcesims} below) was as close to the real observation as possible. \\ 
The objective of these simulations was to see whether we could recover the mass input into the simulation to create a cluster using the physical model, `observed' by AMI and then analysed using the same model. We tried this for the four sets of simulations described below.  \\
For each simulation different noise / canonical radio-source environment realisations (where relevant) were used each time. Due to the large sample size this should not affect any systematic trends seen in the results, and it avoids having to pick a particular realisation to be used in all the simulations. 
%Furthermore the four contributions were generated again for each simulation. This was to avoid the subjective question of `what type of contribution should we add from feature X?', which can only be really addressed by creating many simulations with different `good' or `bad' contributions from each feature, to see exactly how eacFh one affects the parameter estimates. %Consequently in the work presented here, we look at the ability of the physical model to recover parameter estimates from a spectrum of different statistically generated backgrounds over the full sample of clusters with different masses.

% \begin{figure*}%this is relevant for last Section
%   \begin{center}
%   \includegraphics[ width=0.90\linewidth]{figs/M500_AMI_Pl_vs_row_2.ps}
%   \caption{Plot of $M(r_{500})$ vs row number of Table~\ref{tab:results}. 
% The plus shaped markers correspond to $M_{\rm AMI}(r_{500})$ where $z$ was measured spectroscopically, the cirular markers correspond to $M_{\rm \textit{Planck}, \,marg}(r_{500})$ where $z$ was measured spectroscopically, and the star shaped markers correspond to $M_{\rm \textit{Planck}, \,slice}(r_{500})$ where $z$ was measured spectroscopically. The cross shaped markers correspond to $M_{\rm AMI}(r_{500})$ where $z$ was measured photometrically, the diamond markers correspond to $M_{\rm \textit{Planck}, \,marg}(r_{500})$ where $z$ was measured photometrically, and the square markers correspond to $M_{\rm \textit{Planck}, \,slice}(r_{500})$ where $z$ was measured photometrically. }
% \label{graph:m500\textit{Planck}}
%   \end{center}
% \end{figure*}

%%%%%%%%%%%%%%%%%%%%%%%%%%%%%%%%%%%%%%%%%%%%%%%%%%%%%%%%%%%%%%%%%%%%%%%%%%%%%%%

\subsection{Simulations of clusters plus instrumental noise}
\label{subsec:NSNBsims}
\begin{figure}
  \begin{center}
  \includegraphics[ width=0.90\linewidth]{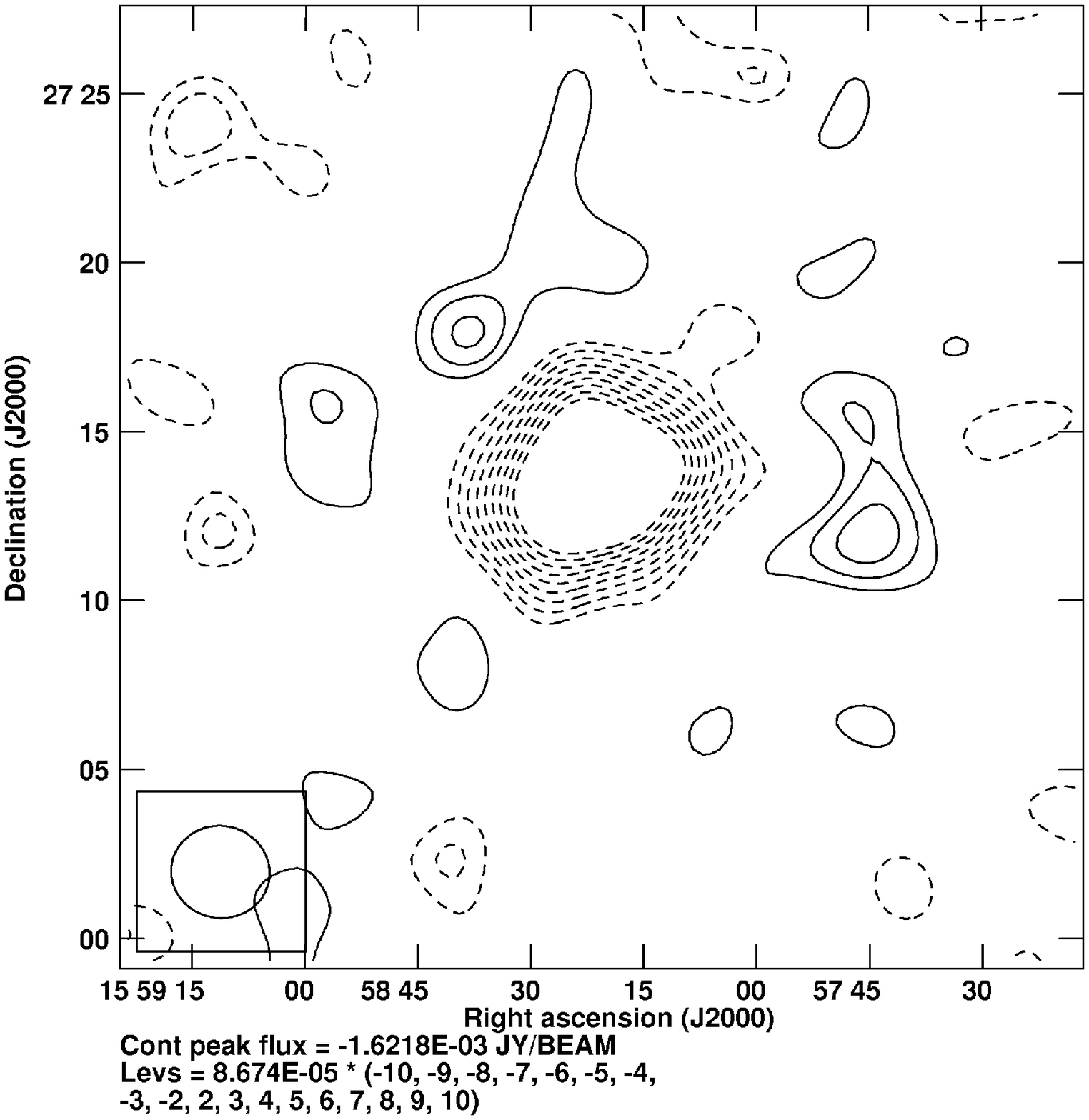}
  \caption{Unsubtracted map produced from simulated AMI data of cluster PSZ2~G044.20+48.66, including instrumental noise.}
  \label{graph:sim_NSNB_map}
  \end{center}
\end{figure}

\begin{figure}
  \begin{center}
  \includegraphics[ width=0.90\linewidth]{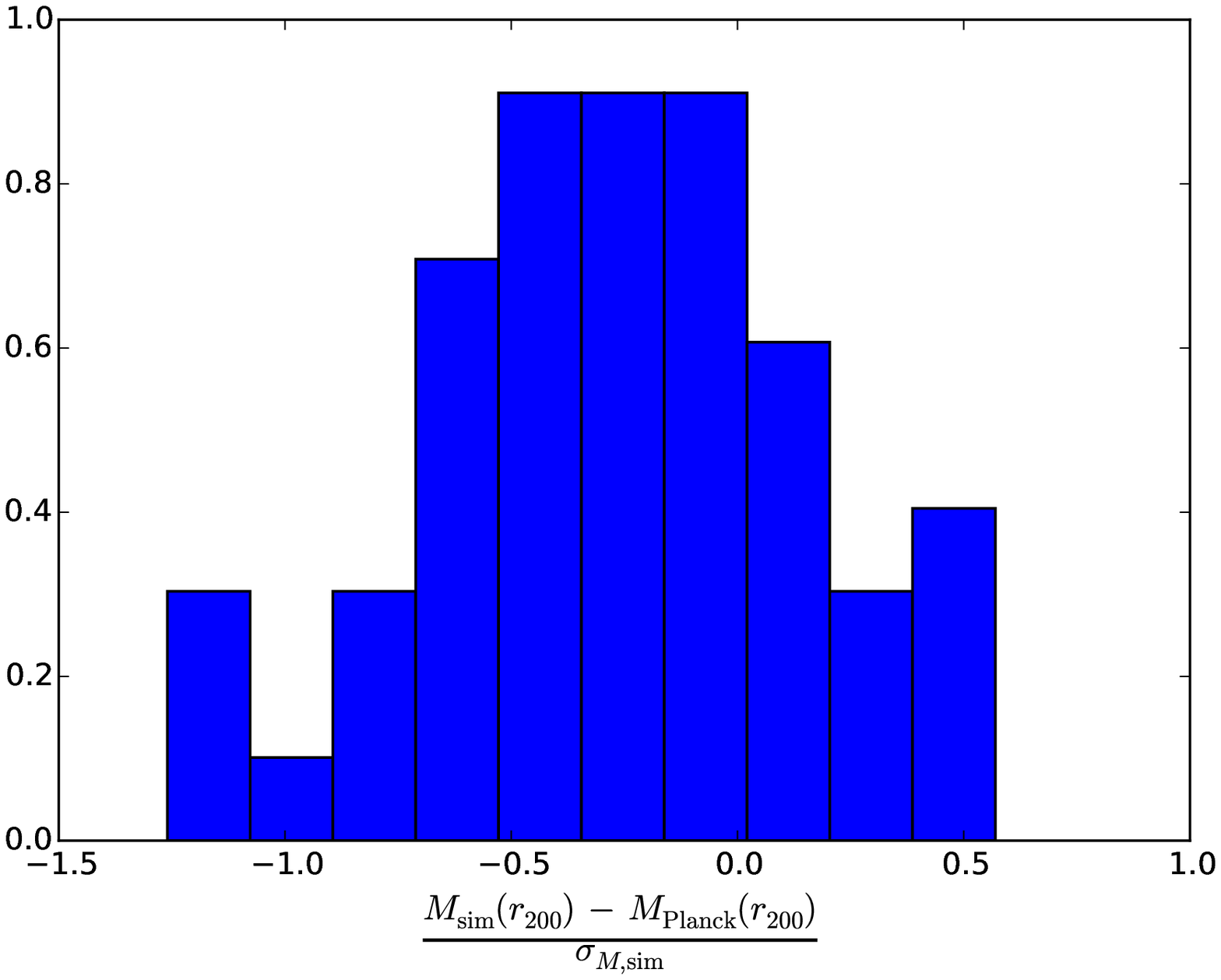}
  \caption{Normalised histogram of the differences between the input and output masses of the AMI simulations including the cluster and instrumental noise only, in units of standard deviations of the output mass.}
\label{graph:simulatednsnb}
  \end{center}
\end{figure}

For each cluster, $M(r_{200})$ was calculated and Gaussian instrumental noise was added to the sky. The RMS of the noise added was $0.7~ \rm{Jy}$ per channel per baseline per second, a value typical of an AMI cluster observation. %Since AMI makes measurements of the instrumental noise when it makes an observation (this is also the case in the simulated observations), its inclusion should not have a large effect on cluster parameter estimates, but was required to make the covariance matrix non-singular. 
Figure~\ref{graph:sim_NSNB_map} shows the map produced from the simulated data of cluster PSZ2~G044.20+48.66 plus this instrumental noise. The mass estimate derived from the Bayesian analysis of this cluster is 0.56 standard deviations above the input value.\\
Figure~\ref{graph:simulatednsnb} shows the difference between the input masses and the ones recovered from running the simulated observations through \textsc{McAdam}, visualised using a histogram. All but three of the clusters lie within one standard deviation of the input mass, and even these clusters (PSZ2~G154.13+40.19, PSZ2~G207.88+81.31 and PSZ2~G213.39+80.59) give an output mass 1.01, 1.26 and 1.08 standard deviations below the input mass. 

%\begin{figure}
%  \begin{center}
%  \includegraphics[ width=0.90\linewidth]{figs/M200_NSNB}
%  \caption{Plot of the differences between the input and output masses of the AMI simulations, in units of standard deviations of the output mass, against row number. This is the case for no radio-source environment, and no background noise (except for instrumental).}
%\label{graph:simulatednsnb}
%  \end{center}
%\end{figure}

%%%%%%%%%%%%%%%%%%%%%%%%%%%%%%%%%%%%%%%%%%%%%%%%%%%%%%%%%%%%%%%%%%%%%%%%%%%%%%%

\subsection{Simulations further adding confusion noise and primordial CMB}
\label{subsec:noisesims}
\begin{figure}
  \begin{center}
  \includegraphics[ width=0.90\linewidth]{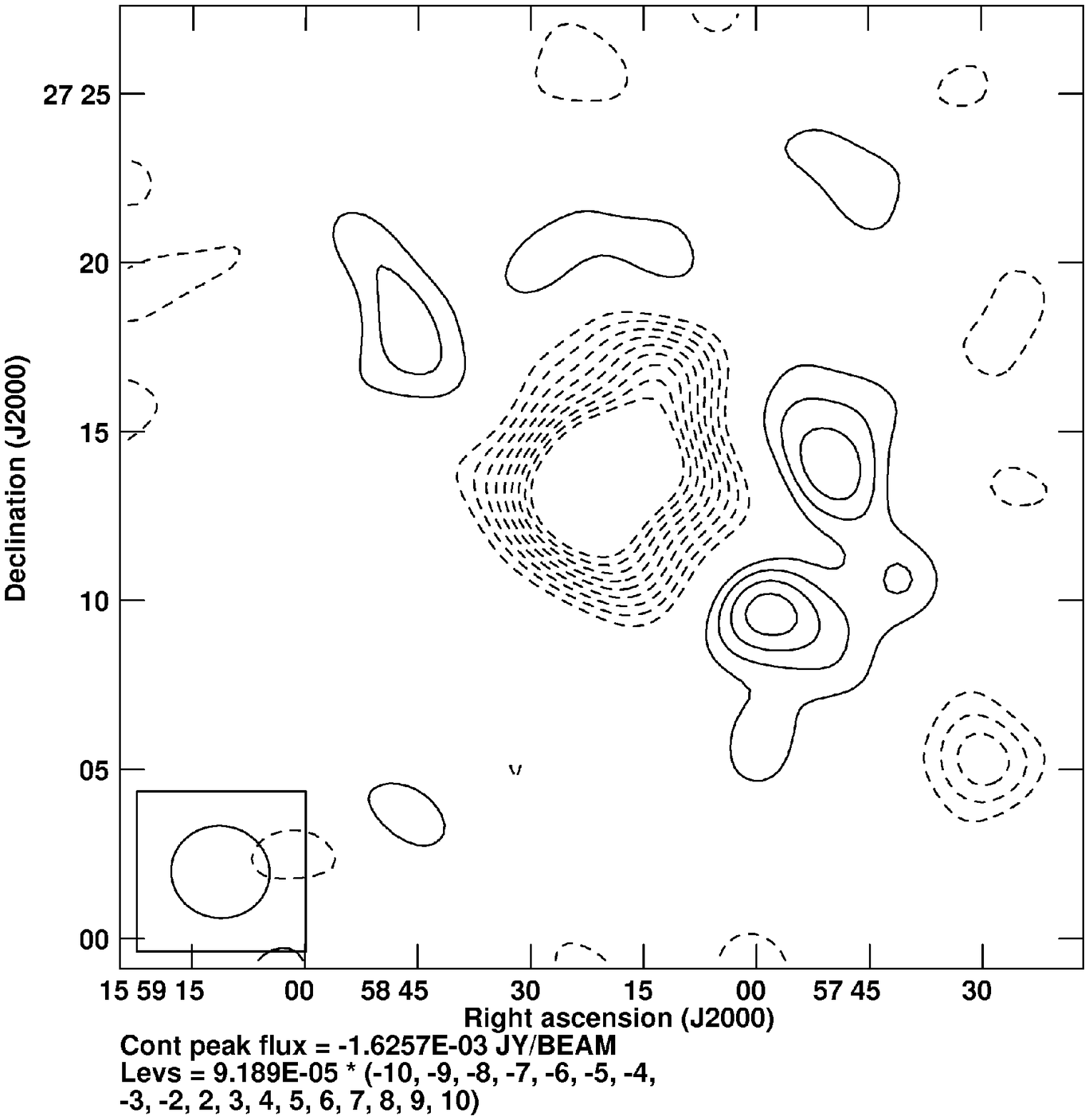}
  \caption{Unsubtracted map produced from simulated AMI data of cluster PSZ2~G044.20+48.66, including instrumental, confusion and CMB noise.}
  \label{graph:sim_NS_map}
  \end{center}
\end{figure}

\begin{figure}
  \begin{center}
  \includegraphics[ width=0.90\linewidth]{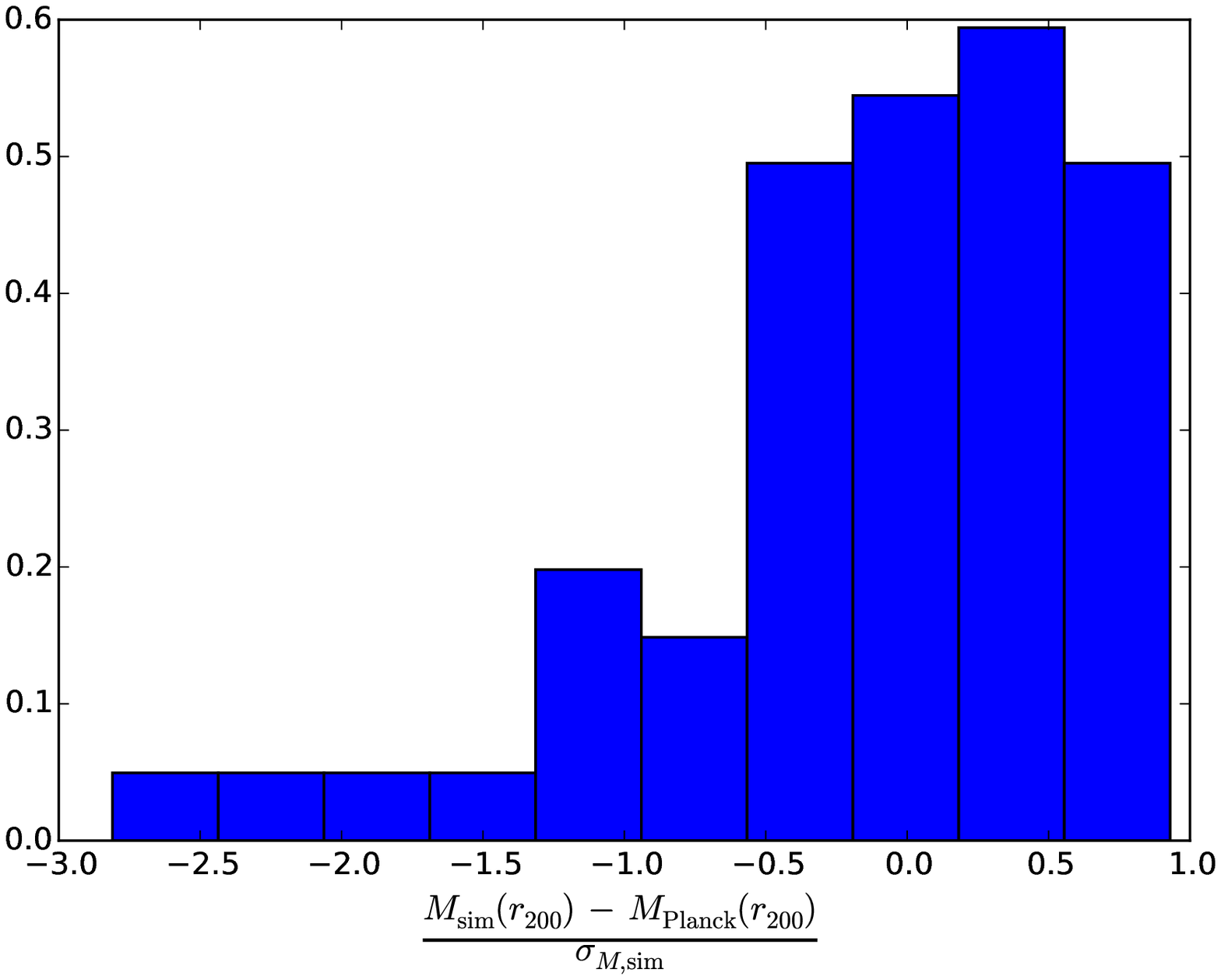}
  \caption{Normalised histogram of the differences between the input and output masses of the AMI simulations, in units of standard deviations of the output mass. This is the case for instrumental, confusion and CMB noise contributions.}
\label{graph:simulatedns}
  \end{center}
\end{figure}

Confusion noise is defined to be the flux from radio-sources below a certain limit (here $S_{\rm{conf}} = 0.3~\rm{mJy}$). In this Section all radio-source realisations only contribute to the confusion noise. However in Sections~\ref{subsec:CSsims} and~\ref{subsec:sourcesims} sources above $S_{\rm{conf}}$ are included. The confusion noise contributions (see e.g. Section~5.3 of FF09) were sampled from the probability density function corresponding to the 10C source counts given in \citet{2011MNRAS.415.2708A}, and placed at positions chosen at random. Similarly, the primordial CMB realisations were sampled from an empirical distribution \citep{2013ApJS..208...19H}, and randomly added to the maps. \\
Figure~\ref{graph:sim_NS_map} shows the map produced from the simulated data of cluster PSZ2~G044.20+48.66, including the three noise contributions. The mass estimate derived from the Bayesian analysis of this cluster is 0.22 standard deviations above the input value.
The differences between output and input masses are shown in Figure~\ref{graph:simulatedns}. This time eight out of the 54 clusters cannot recover the input mass to within one standard deviation. In all eight of these cases, the mass is underestimated with respect to the input value. Five of the outlier values correspond to clusters at low redshift ($z < 0.2$). %Include the following in thesis, but not here. It will also need to be worked on somewhat: One would expect confusion and primordial CMB noise to add bias to the results. Even though the models used to calculate the theoretical covariance matrix for these quantities are the same as the ones used in the simulations, it is impossible for \textsc{McAdam} to model the randomness of the allocation. Similarly for real observations, it is impossible for \textsc{McAdam} to distinguish these noise contributions from the other signal contributors, and so these simulations show how such uncertainty can bias the final results.

%\begin{figure}
%  \begin{center}
%  \includegraphics[ width=0.90\linewidth]{figs/M200_NS}
%  \caption{Plot of the differences between the input and output masses of the AMI simulations, in units of standard deviations of the output mass, against row number. This is the case for no radio-source environment, but with primordial CMB, confusion and instrumental noise contributions added.}
%\label{graph:simulatedns}
%  \end{center}
%\end{figure}

%%%%%%%%%%%%%%%%%%%%%%%%%%%%%%%%%%%%%%%%%%%%%%%%%%%%%%%%%%%%%%%%%%%%%%%%%%%%%%%

\subsection{Simulations further adding a canonical radio-source environment}
\label{subsec:CSsims}
\begin{figure}
  \begin{center}
  \includegraphics[ width=0.90\linewidth]{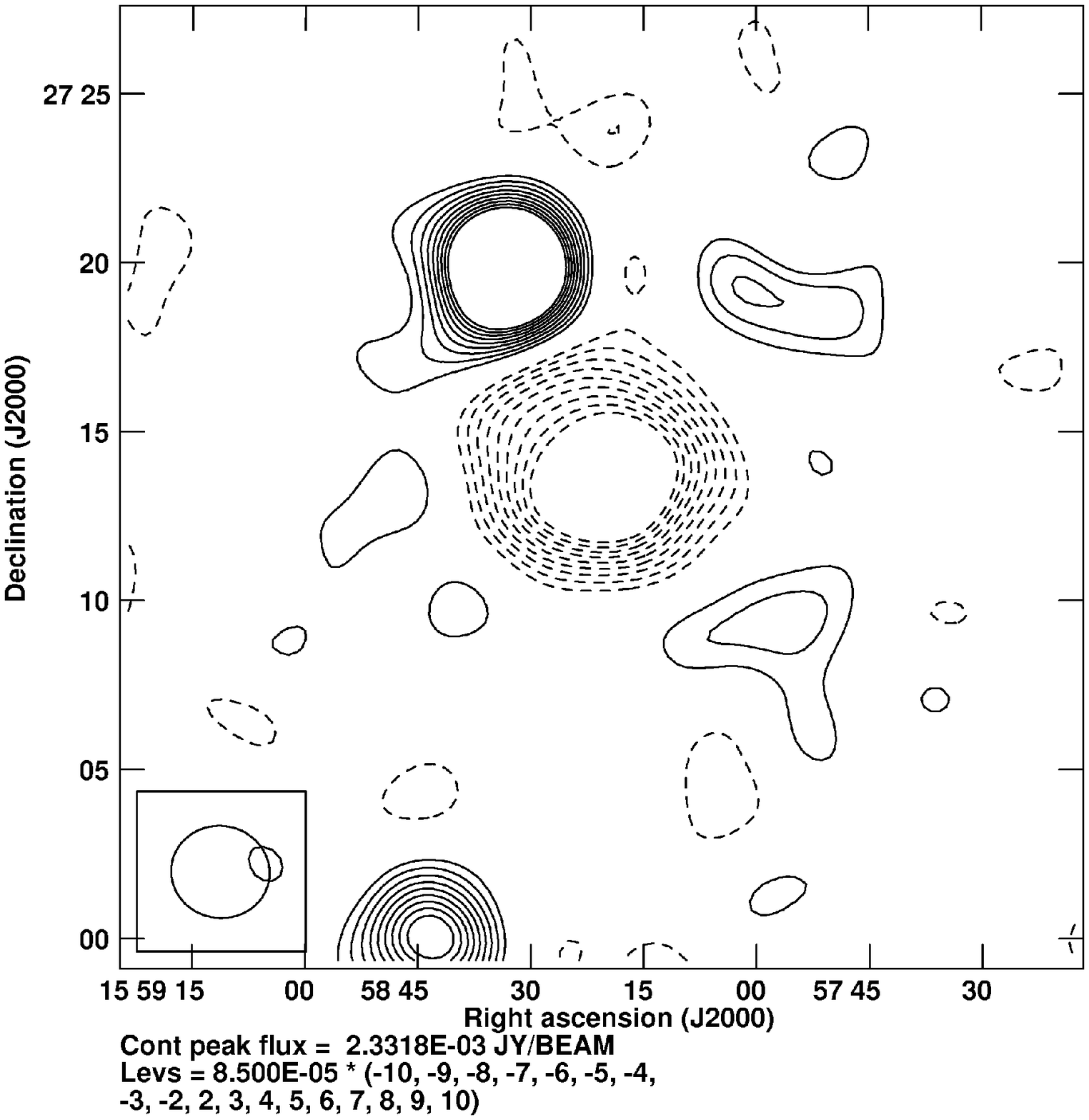}
  \caption{Unsubtracted map produced from simulated AMI data of cluster PSZ2~G044.20+48.66, including a canonical radio-source environment as well as instrumental, confusion and CMB noise.}
  \label{graph:sim_CS_map}
  \end{center}
\end{figure}

\begin{figure}
  \begin{center}
  \includegraphics[ width=0.90\linewidth]{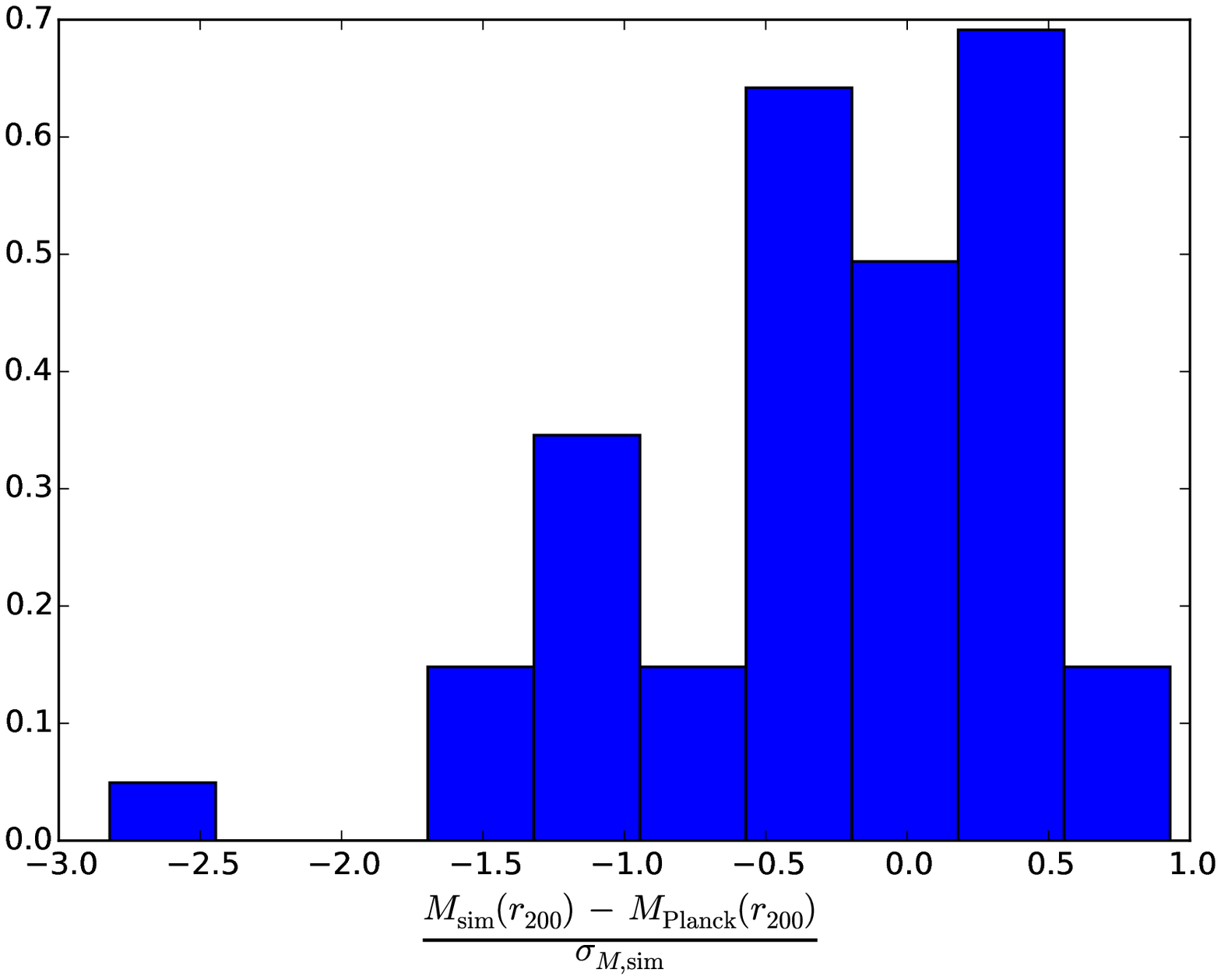}
  \caption{Normalised histogram of the differences between the input and output masses of the AMI simulations, in units of standard deviations of the output mass. This is the case for a canonical radio-source environment as well instrumental, confusion and CMB noise contributions.}
\label{graph:simulatedcs}
  \end{center}
\end{figure}

The third set of simulations included recognised radio-sources, which formed a canonical radio-source environment. They were created in the same way as with the confusion noise described above, but with higher flux limits so that in reality, the LA would have been able to recognise them. The upper flux limit was set to $25~\rm{mJy}$.\\ %The `measured' flux values from the LA were set to be the values used in the simulation, meaning that the corresponding priors in \textsc{McAdam} were either delta functions or Gaussians centred on the actual flux values of the sources. \\
Figure~\ref{graph:sim_CS_map} shows the map produced from the simulated data of cluster PSZ2~G044.20+48.66, including a canonical source environment and background noise. The mass estimate derived from the Bayesian analysis of this cluster is 0.51 standard deviations below the input value.
Figure~\ref{graph:simulatedcs} shows that the canonical radio-source environment have little effect on the mass estimation relative to Section~\ref{subsec:noisesims}, as there are still 8 clusters which give mass estimates greater than one standard deviation away from the input value. %This suggests that `ideal' radio-sources, that is, ones which can have their positions and fluxes measured perfectly and have circular symmetry in the sky plane, have little effect on cluster parameter estimates. 
Note that in this case, the outliers occurred across the entire range of redshifts, which suggests that in Section~\ref{subsec:noisesims} the low redshift trend was just a coincidence. %This seems plausible, since the CMB and confusion contributions were randomly generated for each cluster, and so 5 low redshift simulations could have been subject to `nasty' contributions by chance, which were difficult for \textsc{McAdam} to model.

%\begin{figure}
%  \begin{center}
%  \includegraphics[ width=0.90\linewidth]{figs/M200_CS}
%  \caption{Plot of the differences between the input and output masses of the AMI simulations, in units of standard deviations of the output mass, against row number. This is the case for a canonical radio-source environment, with primordial CMB, confusion and instrumental noise contributions added.}
%\label{graph:simulatedcs}
%  \end{center}
%\end{figure}

%%%%%%%%%%%%%%%%%%%%%%%%%%%%%%%%%%%%%%%%%%%%%%%%%%%%%%%%%%%%%%%%%%%%%%%%%%%%%%%

\subsection{Simulations with LA observed radio-source environment plus instrumental, confusion and CMB noise}
\label{subsec:sourcesims}
\begin{figure*}
  \begin{center}
  \includegraphics[ width=0.45\linewidth]{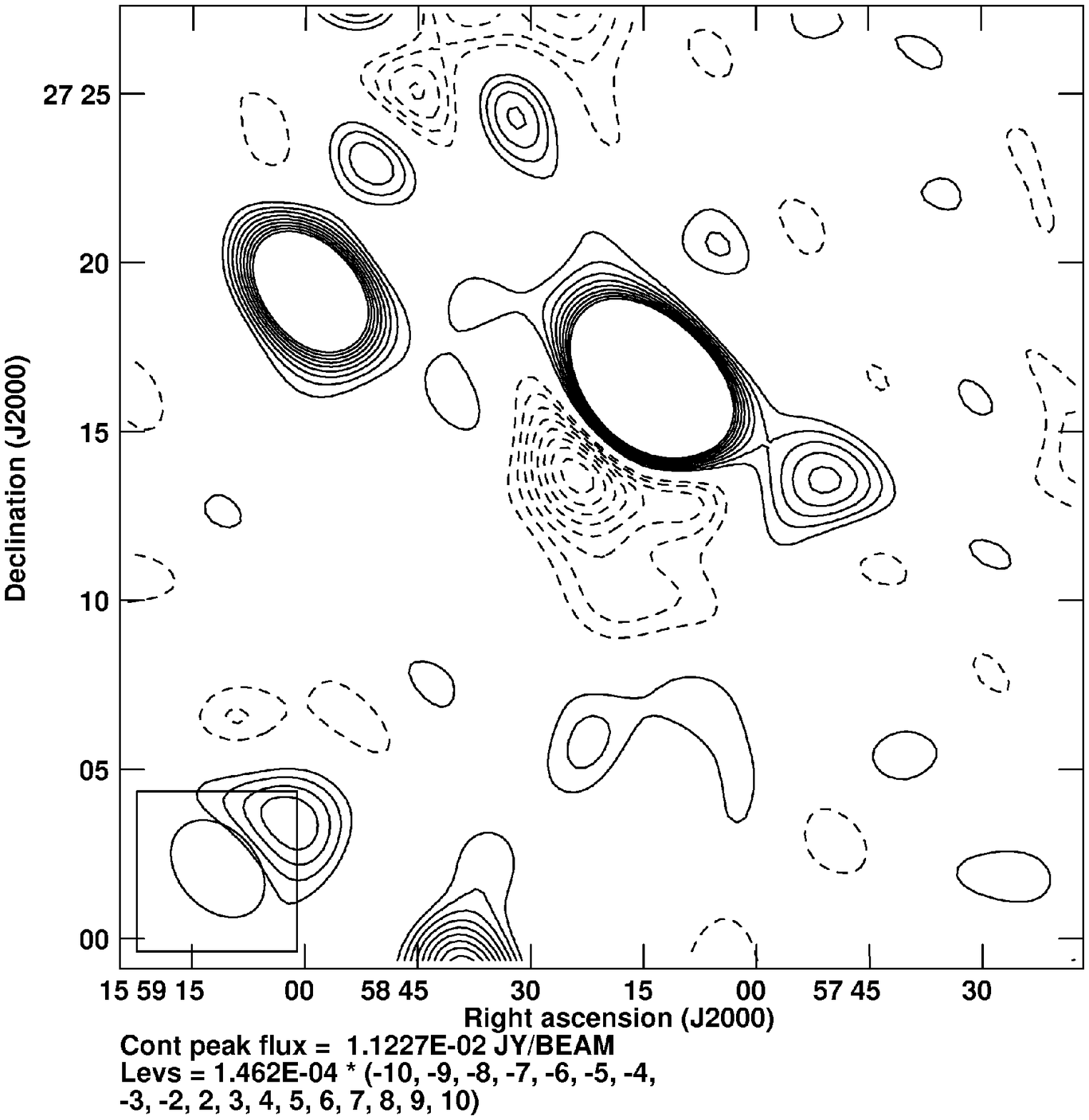}
  \includegraphics[ width=0.45\linewidth]{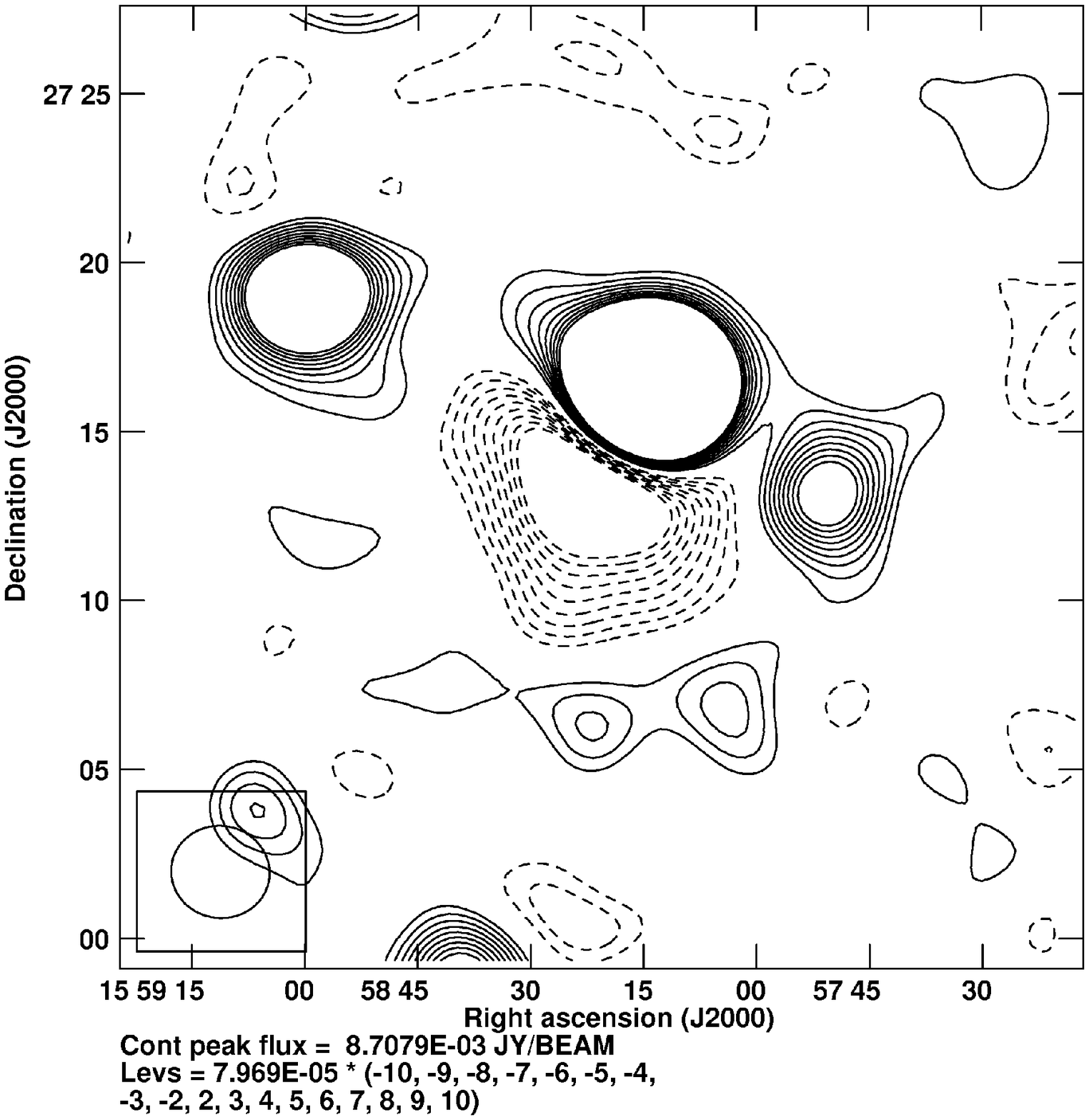}
  \medskip
  \centerline{(a) \hskip 0.45\linewidth (b)}
  \caption{(a) Unsubtracted map produced from real AMI data of cluster PSZ2~G044.20+48.66. (b) Unsubtracted map produced from simulated AMI data of PSZ2~G044.20+48.66, including the real source environment (as measured by the LA) as well as instrumental, confusion and CMB noise. The peak flux in the simulation has been underestimated relative to the real observation by $\approx 25\%$. This could be due to the source sitting on a negative decrement caused by background noise, or it could be from the cluster decrement.}
  \label{graph:sim_rs_map}
  \end{center}
\end{figure*}

\begin{figure}%this Figure is relevant to the previous Section, but is included here to stop the formatting messing up.
  \begin{center}
  \includegraphics[ width=0.90\linewidth]{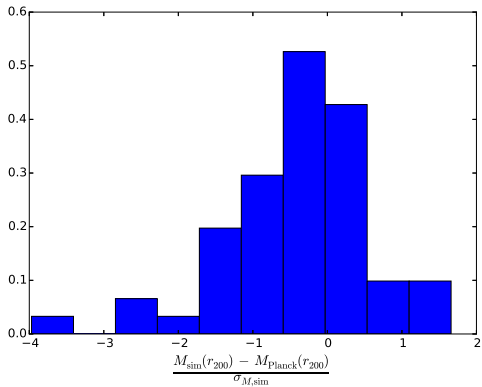}
  \caption{Normalised histogram of the differences between the input and output masses of the AMI simulations, in units of standard deviations of the output mass. This is the case for the real radio-source environment as measured by the LA, with instrumental, confusion and CMB noise contributions.}
\label{graph:simulatedrs}
  \end{center}
\end{figure}

The final set of simulations included the radio-source environment measured by the LA during the real observation for each cluster. These are only estimates of the actual source environments, and are only as reliable as the LA's ability to measure them. %Nevertheless, the fact that the values used for the priors in \textsc{McAdam} are the same as the input in the simulations  suggests that they shouldn't cause any large biases. \\
Figure~\ref{graph:sim_rs_map} shows the maps produced from the real and simulated data of cluster PSZ2~G044.20+48.66. The mass estimate derived from the Bayesian analysis of the simulated dataset is just 0.08 standard deviations above the input value. \\
Figure~\ref{graph:simulatedrs} shows that including the LA observed radio-source environment has a large effect on the results, as this time there are 16 clusters which are more than one standard deviation away from the input mass. Furthermore, three of these overestimated the mass relative to the input, the first time we have seen this occur in any of the simulations. 
%This suggests that for real source environments, even when the positions and fluxes are known, biases can still occur. This 
A possible source of bias could be due to for example, the empirical prior on the spectral index incorrectly modelling some radio-sources. %For real observations, the digital correlator upgrade should improve the LA's ability to measure the spectral index of a source. For higher flux sources, the LA may be able to measure the spectral index accurately enough so that the empirical Waldram prior is no longer necessary.
Another source of bias could be the position of a source relative to the cluster, and the magnitude of the source flux. For example, if a high flux radio-source is close to the centre of the galaxy cluster, then even a slight discrepancy between the real and the modelled values for the source could have a large effect on the cluster parameter estimates.

We now compare these results to the simulations in YP15 (which concluded that the underestimation of the simulation input values could be due to deviation from the `universal' profile, see Figure~23a in the paper). The results of the large cluster simulations (total integrated Comptonisation parameter $= 7 \times 10^3$~arcmin$^2$ and $\theta_{\rm p} = 7.4$~arcmin) in YP15 seem biased low at a more significant level than those in Figure~\ref{graph:simulatedrs}, as in the former case less than half of the clusters recover the true value within two standard deviations. For the smaller clusters however, YP15 found a slight upward bias in the simulation results, but this is probably smaller in magnitude than the bias found in this Section. \\

%\begin{figure}
%  \begin{center}
%  \includegraphics[ width=0.90\linewidth]{figs/M200_RS}
%  \caption{Plot of the differences between the input and output masses of the AMI simulations, in units of standard deviations of the output mass, against row number. This is the case for the real radio-source environment as measured by the LA, with primordial CMB, confusion and instrumental noise contributions added.}
%\label{graph:simulatedrs}
%  \end{center}
%\end{figure}

\subsection{Statistics of results of real and simulated data}
\label{subsec:simsstats}

Looking at the histograms produced in Sections~\ref{subsec:NSNBsims},~\ref{subsec:noisesims},~\ref{subsec:CSsims}, and~\ref{subsec:sourcesims}, in the last three cases it is apparent that there is a negative skew in the data, i.e. the output masses are systematically low relative to the input masses. The skews calculated from the samples associated with the four histograms are $-0.17$, $-1.30$, $-0.91$, and $-0.96$ respectively in units of standard deviations of the output mass. This suggests that the inclusion of confusion and CMB noise bias the AMI cluster mass estimates. 
We also calculate the median values associated with the histograms, and compare them with the medians corresponding to the real AMI and PSZ2 masses given in Figure~\ref{graph:m500planck}. The median values for the four histograms are $-0.24$, $0.09$, $-0.27$ and $-0.34$ respectively in units of standard deviations of the output mass. For the real data the median values for $(M_{\rm AMI}(r_{500}) - M_{\rm Pl,\, marg}(r_{500})) / \sigma_{\rm AMI}$ and $(M_{\rm AMI}(r_{500}) - M_{\rm Pl,\, slice}(r_{500})) / \sigma_{\rm AMI}$ are $-1.57$ and $-0.56$. It makes sense to compare the second of these real data values with those obtained from the simulations, as it was $M_{\rm Pl,\, slice}(r_{500})$ which was used to derive the input masses. The fact that the median from the real data is greater in magnitude than the values from the simulations implies in general, our simulations can recover their input values with better agreement than that obtained between real AMI estimates and those obtained from \textit{Planck} data using the slicing function methodology. This seems plausible as you would expect that inferring results from data which was created using the same model used in the inference would be more accurate than results from data taken from two different telescopes, which use different models in their inference. 
Furthermore the simulation medians tell us that AMI is capable of inferring the masses derived with the slicing methodology, if the cluster is created using the model used in the inference and assuming there are no large discrepancies between the real and simulated AMI observations.

%%%%%%%%%%%%%%%%%%%%%%%%%%%%%%%%%%%%%%%%%%%%%%%%%%%%%%%%%%%%%%%%%%%%%%%%%%%%%%%
%%%%%%%%%%%%%%%%%%%%%%%%%%%%%%%%%%%%%%%%%%%%%%%%%%%%%%%%%%%%%%%%%%%%%%%%%%%%%%%
%%%%%%%%%%%%%%%%%%%%%%%%%%%%%%%%%%%%%%%%%%%%%%%%%%%%%%%%%%%%%%%%%%%%%%%%%%%%%%%
  
\section{Conclusions}
\label{sec:summary}

We have made observations of galaxy clusters detected by the \textit{Planck} space telescope, with the Arcminute Microkelvin Imager (AMI) radio interferometer system in order to compare mass estimates obtained from their data. We then analysed this data using a physical model based on the one described in \citet{2012MNRAS.423.1534O}, following largely the data analysis method outlined in \citet{2009MNRAS.398.2049F}. This allowed us to derive physical parameter estimates for each cluster, in particular the total mass out to a given radius. We have also calculated two mass estimates for each cluster from \textit{Planck}'s PowellSnakes detection algorithm \citep{2012MNRAS.427.1384C} data following \citet{2016A&A...594A..27P} (PSZ2). We found the following.
% and cluster parameter scaling relations which rely on external X-ray measurements

\begin{itemize}
\item For the AMI mass estimates of \textit{Planck} selected clusters there is generally a steeping in the mass of galaxy clusters as a function of redshift, which flattens out at around $z \approx 0.5$.
%\item The minimum AMI mass estimates of \textit{Planck} selected clusters increases with redshift up to $z \approx 0.5$, while the maximum mass appears roughly independent of $z$.
\item AMI $M(r_{500})$ estimates are within one combined standard deviation of the PSZ2 slicing function mass estimates for 31 out of the final sample of 54 clusters. However, the AMI masses are lower than both PSZ2 estimates for 37 out of the 54 cluster sample.
\end{itemize}

To investigate further the possible biasing of AMI mass estimates, we created simulations of AMI data with input mass values from the PSZ2 slicing methodology. We considered four different cases for the simulations: 1) galaxy cluster plus instrumental noise; 2) galaxy cluster plus instrumental plus confusion and CMB noise; 3) galaxy cluster plus instrumental, confusion and CMB noise, plus a randomly positioned radio-source environment; 4) galaxy cluster plus instrumental, confusion and CMB noise, plus the radio-source environment detected by the LA in the real observations. These simulated datasets were analysed in the same way as the real datasets, and we found the following.

\begin{itemize}
\item For case 1), the physical model recovered the input mass to within one standard deviation for 51 of the 54 clusters. The three which did not give an underestimate relative to the masses input to the simulation.
\item For case 2), eight of the simulations gave results which were more than one standard deviation lower than the input values. This highlights the effect of incorporating the noise sources into the error covariance matrix rather than trying to model the associated signals explicitly. %This highlights the difficulty in trying to model noise contributions with an element of randomness. In the simulated case this comes in the form of the physical location of the noise being completely random, and its magnitude being treated as random variables drawn from empirical probability distributions. In the real observations this effect is amplified as the noise does not even follow the empirical relations exactly.
\item Case 3) shows similar results to case 2), which implies that `ideal' radio-sources placed randomly in the sky have little effect on cluster mass estimates.% if the sources are well identified during the observation.
\item However in case 4) with real source environments, 16 simulations did not recover the input mass to within one standard deviation. This suggests that real radio-source environments, which can include sources with high flux values, and often sources which are located very close to the cluster centre, introduce biases in the cluster mass estimates. In real observations there are also additional issues (the sources are not `ideal'), such as sources being extended and emission not being circularly symmetric on the sky.%, and of course by the fact that their positions, fluxes and spectral indices are never known exactly.
\item Cases 2), 3) and 4) give distributions of output $-$ input mass which are negatively skewed. Thus AMI mass estimates are expected to be systematically lower than the PSZ2 slicing methodology values. 
\item Compared to the results of simulations of large clusters carried out in \citet{2015A&A...580A..95P}, which test the robustness of the `universal' pressure profile, the case 4) bias appears relatively small in magnitude, and in the same direction (downward). When comparing the case 4) results with the small cluster simulations of \citet{2015A&A...580A..95P}, the latter shows a relatively small bias in the opposite direction.
\item The median values of the distributions of output $-$ input mass of the simulations in each of the four cases are smaller in magnitude than those obtained from comparing AMI and PSZ2 estimates from real data. This is expected as we are using the same model to simulate and analyse the clusters in all four cases. %suggesting that AMI is capable of obtaining the masses inferred in the PSZ2 slicing function methodology, given the clusters are well described by the physical model used in the Bayesian analysis.
\item The simulated and real data medians also indicate that while the simulations have shown that AMI mass estimates are systematically low, this does not fully accommodate for the discrepancies in the results obtained from the real data. This suggests that there is a systematic difference between the AMI and \textit{Planck} data and / or the cluster models used to determine the mass estimates (which generally leads to PSZ2 estimates being higher than those obtained from AMI data).
\end{itemize}
%The results presented here provide an opportunity for a number of different follow-ups: \\
In a forthcoming paper \citep{2018arXiv180501968J}, comparison of the `observational' parameters (i.e. the integrated Comptonisation parameter $Y$ and the angular radius $\theta$) obtained from AMI data will be analysed. Furthermore, in \citet{2019MNRAS.489.3135J} and \citet{2019arXiv190900029J}, different dark matter and pressure models will be considered, and in \citet{2019MNRAS.486.2116P}, Bayesian analysis will be performed on joint AMI-\textit{Planck} datasets.
% Furthermore a comparison of the values obtained using \textit{Planck} methodology with AMI observed data (as in YP15) will be made. The `detection ratios' given by the Bayes factor defined in Section~5.5 of FF09 will also be compared between the various datasets and models. \\
%\textit{Planck} detected clusters will be re-observed with AMI. Now that the array uses digital correlators, it is able to observe at lower declination, and is able to make measurements at a higher degree of accuracy. This will affect the observations of the clusters themselves with the SA, and also the measurements of the external radio-source environment made by the LA. It will be interesting to see how the upgrade affects the parameter estimates derived from AMI data, and whether well-constrained posterior distributions can be produced for clusters which were not in this analysis. \\

%There are also plans to formulate a new physical model, using the dark matter profile derived by \citet{1965TrAlm...5...87E}, instead of the NFW profile given by equation~\ref{eqn:nfw}.

%%%%%%%%%%%%%%%%%%%%%%%%%%%%%%%%%%%%%%%%%%%%%%%%%%%%%%%%%%%%%%%%%%%%%%%%%%%%%%%
%%%%%%%%%%%%%%%%%%%%%%%%%%%%%%%%%%%%%%%%%%%%%%%%%%%%%%%%%%%%%%%%%%%%%%%%%%%%%%%
%%%%%%%%%%%%%%%%%%%%%%%%%%%%%%%%%%%%%%%%%%%%%%%%%%%%%%%%%%%%%%%%%%%%%%%%%%%%%%%

\section*{Acknowledgements}
This work was performed using the Darwin Supercomputer of the University of Cambridge High Performance Computing (HPC) Service (\url{http://www.hpc.cam.ac.uk/}), provided by Dell Inc. using Strategic Research Infrastructure Funding from the Higher Education Funding Council for England and funding from the Science and Technology Facilities Council. The authors would like to thank Stuart Rankin from HPC and Greg Willatt and David Titterington from Cavendish Astrophysics for computing assistance. They would also like to thank Dave Green for his invaluable help using \LaTeX.
Kamran Javid acknowledges an STFC studentship. Yvette Perrott acknowledges support from a Trinity College Junior Research Fellowship.

%%%%%%%%%%%%%%%%%%%%%%%%%%%%%%%%%%%%%%%%%%%%%%%%%%%%%%%%%%%%%%%%%%%%%%%%%%%%%%%
%%%%%%%%%%%%%%%%%%%%%%%%%%%%%%%%%%%%%%%%%%%%%%%%%%%%%%%%%%%%%%%%%%%%%%%%%%%%%%%
%%%%%%%%%%%%%%%%%%%%%%%%%%%%%%%%%%%%%%%%%%%%%%%%%%%%%%%%%%%%%%%%%%%%%%%%%%%%%%%

\setlength{\bibsep}{0pt}            % vertical spacing between references
\renewcommand{\bibname}{References} % instead of "Bibliography"

%%%%%%%%%%%%%%%%%%%%%%%%%%%%%%%%%%%%%%%%%%%%%%%%%%%%%%%%%%%%%%%%%%%%%%%%%%%%%%%
%%%%%%%%%%%%%%%%%%%%%%%%%%%%%%%%%%%%%%%%%%%%%%%%%%%%%%%%%%%%%%%%%%%%%%%%%%%%%%%
%%%%%%%%%%%%%%%%%%%%%%%%%%%%%%%%%%%%%%%%%%%%%%%%%%%%%%%%%%%%%%%%%%%%%%%%%%%%%%%

%\newgeometry{margin=1cm} %old location of results table
%\onecolumn
%\begin{landscape}
%\appendix
%\section{Results table}\label{sec:results_Table}
%\begin{center}
%\begin{longtable}{llllllllllll}

% Don't change these lines
\bsp	% typesetting comment
\label{lastpage}
%\end{landscape}
%\restoregeometry
\end{document}